\documentclass[10pt]{article}
\usepackage{graphicx}
\usepackage{amsmath}
\usepackage{amssymb}
\usepackage{caption2}
\begin{document}
\begin{center}
{\large {\bf \sc{  $B-S$ transition form-factors with the light-cone   QCD sum rules }}} \\[2mm]
Zhi-Gang Wang \footnote{E-mail,zgwang@aliyun.com.  }    \\
 Department of Physics, North China Electric Power University,
Baoding 071003, P. R. China
\end{center}

\begin{abstract}
In the article,  we assume  the two scalar nonet mesons below and above 1 GeV are all $\bar{q}q$ states,
  in case I, the scalar mesons below 1 GeV are the ground states, in case II, the scalar mesons above 1 GeV are the ground states.
   We calculate the $B-S$ form-factors by taking into account the perturbative  ${\mathcal{O}}(\alpha_s)$
 corrections to the twist-2 terms using the light-cone   QCD sum rules and fit the numerical values of the form-factors into the single-pole forms,
 which have many phenomenological applications.
\end{abstract}



\section{Introduction}

The underlying structures of the scalar mesons are not well established theoretically,
there are many candidates  with the quantum numbers $J^{PC}=0^{++}$ below $2\,\rm{GeV}$, which
cannot be accommodated in one $\bar{q}q$ nonet.
 Roughly speaking, they can be calcified into  two nonets,
the nonet  $\{f_0(600)$, $a_0(980)$, $\kappa(800)$, $f_0(980) \}$ below 1 GeV  and
the nonet  $\{f_0(1370)$, $a_0(1450)$, $K^*_0(1430)$, $f_0(1500) \}$ above 1 GeV.
A prospective picture  suggests that the scalar mesons above 1 GeV can be assign to be a conventional $ \bar{q}q$
nonet with some possible glue  components, while the scalar mesons below 1 GeV form
 an exotic  $[qq]_{\bar{3}}[\bar{q}\bar{q}]_{3}$ nonet  with  substantial   mixings with the $q\bar{q}$ states, meson-meson states and glueballs \cite{ReviewAmsler}.
 In the  hadronic dressing mechanism, the scalar mesons below 1 GeV  have small $ \bar{q}q$
cores of typical $ \bar{q}q$ meson size,  strong couplings to the
intermediate hadronic states   enrich the pure
$ \bar{q}q$ states with other components and spend part  of their lifetime as virtual meson-meson states \cite{HDress}.
 All the existing pictures  have both advantages and shortcomings in one way or another.
 In this article, we assume that the scalar mesons are all $\bar{q}q$ states,
  in case I, the scalar mesons below 1 GeV are the ground states, in case II, the scalar mesons above 1 GeV are the ground states; and study the $B-S$ transition form-factors   with the light-cone QCD sum rules (LCSR) \cite{LCSR,LCSR-CFS}.

The transition form-factors in the semi-leptonic decays not only depend on the dynamics of strong interactions
among the quarks in the initial and final mesons, but also depend  on the structures of the involved mesons. They are highly nonperturbative quantities.
In the region small-recoil, where the   momentum transfer squared  $q^2$ is large, the form-factors are dominated by the soft dynamics, while in the large-recoil region,  where $q^2 \to 0$,  the form-factors are dominated by the short-distance dynamics.

In the light-cone QCD sum rules, we carry out the operator product expansion near
the light-cone $x^2\approx0$ in stead of the short distance $x\approx0$, the nonperturbative hadronic
matrix elements are parameterized by the light-cone distribution amplitudes (LCDAs) of
increasing twist instead of the vacuum condensates \cite{LCSR,LCSR-CFS}. Based on the quark-hadron duality,
we can obtain copious information about the hadronic parameters at the phenomenological
side.  The transition form-factors from the LCSR not only have an estimable region of $q^2$, but also
 embody as many long-distance effects as possible involved in the decaying processes.
 The operator product expansion  is valid at small and intermediate momentum transfer squared
$q^2$, $0\leq q^2 \leq(m_b-m_S)^2-2(m_b-m_S)\chi$, where the $\chi$ is a typical hadronic scale of roughly 500 MeV and independent of the $b$-quark mass $m_b$.

On the other hand, the semi-leptonic $B$-decays  are excellent subjects in studying  the CKM matrix elements
and CP violations. We can use both the exclusive and inclusive $b \to u$ transitions to study the
CKM matrix element $V_{ub}$. Furthermore, the processes induced by the flavor-changing neutral currents $b \to s(d)$
  provide the most sensitive and
 stringent  test for the standard model at one-loop level and can put
powerful constraints on the new physics models, as they are forbidden at the tree-level in the standard model \cite{NPhysics}. So reliable calculations of the transition form-factors are needed to make robust predictions.

The $B-S$ transition form-factors have been calculated with the LCSR \cite{WYM08,WW10,SYJ11,WXG13}, the three-point QCD sum rules \cite{YMZ06,Alv07,GK09},  the perturbative QCD \cite{Li09}, the light front quark model \cite{LFQM}, etc.
In Refs.\cite{WYM08,WW10,SYJ11,WXG13}, the $B- S$ form-factors are calculated with the LCSR in the leading order approximation, the perturbative ${\mathcal{O}}(\alpha_s)$ corrections  are neglected. Due to the  lengthy calculations,
only perturbative  ${\mathcal{O}}(\alpha_s)$
 corrections to the twist-2 and twist-3 terms in the LCSR for the  $B - P,\, V$ form-factors \cite{Bpi,Bpi-3,BV}
and  perturbative ${\mathcal{O}}(\alpha_s^2)$
 corrections to the twist-2 terms  in the LCSR for the  $B -\pi$  form-factor \cite{B-pi-two-loop} are studied up to now, where the $P$ and $V$ denote the light pseudoscalar and vector mesons, respectively.
 In this article, we study the $B-S$ form-factors by taking into account the perturbative  ${\mathcal{O}}(\alpha_s)$
 corrections to the twist-2 terms using the LCSR.

The article is arranged as follows:  we derive the LCSR for
the $B-S$ form-factors by including the   perturbative  ${\mathcal{O}}(\alpha_s)$
 corrections to the twist-2 terms  in Sect.2;
in Sect.3, we present the numerical results and discussions; and Sect.4 is reserved for our
conclusions.

\section{Light-cone QCD sum rules for  the form-factors }
In the following, we write down  the two-point correlation functions
$\Pi^{i}_{\mu}(p,q)$  in the LCSR,
\begin{eqnarray}
\Pi^i_{\mu}(p,q)&=&i\int d^4x e^{iq \cdot x}  \langle S(p)|T\left\{J^i_{\mu}(x)J_{5}(0)\right\}|0\rangle\, , \\
J^A_{\mu}(x)&=& \overline{q}(x) \gamma_{\mu}\gamma_5 b(x) \, , \nonumber\\
J^B_{\mu}(x)&=& \overline{q}(x) \sigma_{\mu\nu}\gamma_5q^\nu b(x) \, , \nonumber\\
J_{ 5}(0)&=& m_b\overline{b}(0)i\gamma_5q^{\prime}(0) \, , \nonumber
\end{eqnarray}
where $i=A,B$, $q,q^\prime=u,d,s$, the pseudoscalar currents $J_{ 5}(x)$  interpolate the $B_{q^\prime}$ mesons, and the axial-vector  currents $J^i_\mu(x)$ induce  the $B\to S$ transitions.

We can insert  a complete set of intermediate hadronic states with
the same quantum numbers as the current operators  $J_{ 5}(0)$  into the
correlation functions $\Pi^{i}_{\mu}(p,q)$  to obtain the hadronic representation
\cite{SVZ79,Reinders85}. After isolating the ground state
contributions from the $B_{q^\prime}$ mesons, we get the following result,
\begin{eqnarray}
\Pi^A_{\mu}(p,q)&=&\frac{f_{B_{q^\prime}}m_{B_{q^\prime}}^2 \langle S(p)|J^A_{\mu}(0)|B_{q^\prime}(p+q)\rangle
}{m_{B_{q^\prime}}^2-(p+q)^2}+\cdots\, ,\nonumber\\
&=&-\frac{f_{B_{q^\prime}}m_{B_{q^\prime}}^2
}{m_{B_{q^\prime}}^2-(p+q)^2}\left\{2iF_{+}(q^2)p_\mu +i\left[ F_{+}(q^2)+F_{-}(q^2)\right]q_\mu \right\}+\cdots\, , \\
\Pi^B_{\mu}(p,q)&=&\frac{f_{B_{q^\prime}}m_{B_{q^\prime}}^2 \langle S(p)|J^B_{\mu}(0)|B_{q^\prime}(p+q)\rangle
}{m_{B_{q^\prime}}^2-(p+q)^2}+\cdots\, ,\nonumber\\
&=&-\frac{2f_{B_{q^\prime}}m_{B_{q^\prime}}^2
}{\left(m_{B_{q^\prime}}+m_S\right)\left(m_{B_{q^\prime}}^2-(p+q)^2\right)}\left\{F_{T}(q^2)\left(q^2p_\mu- q\cdot p q_\mu \right)\right\}+\cdots\, ,
\end{eqnarray}
where the  decay constants $f_{B_{q^\prime}}$ and the form-factors $F_{+}(q^2)$, $F_{-}(q^2)$ and $F_{T}(q^2)$   are defined by
\begin{eqnarray}
\langle 0|J_{5}(0)|B_{q^\prime}(p+q)\rangle&=&f_{B_{q^\prime}}m_{B_{q^\prime}}^2  \, , \nonumber\\
\langle S(p)|J^A_{\mu}(0)|B_{q^\prime}(p+q)\rangle&=&-2iF_{+}(q^2)p_\mu -i\left[ F_{+}(q^2)+F_{-}(q^2)\right]q_\mu\, , \nonumber\\
\langle S(p)|J^B_{\mu}(0)|B_{q^\prime}(p+q)\rangle&=&-\frac{2F_T(q^2)}{m_{B_{q^\prime}}+m_S}\left(q^2p_\mu- q\cdot p q_\mu \right)\, .
\end{eqnarray}
We can also  parameterize the form-factors into another form,
\begin{eqnarray}
\langle S(p)|J^A_{\mu}(0)|B_{q^\prime}(p+q)\rangle&=&-i\left[F_{1}(q^2)\left(P_\mu-\frac{m^2_{B_{q^\prime}}-m_S^2}{q^2}q_\mu\right)
+F_{0}(q^2)\frac{m^2_{B_{q^\prime}}-m_S^2}{q^2}q_\mu\right]\, ,
\end{eqnarray}
where $P=2p+q$ and
\begin{eqnarray}
F_{1}(q^2)&=&F_{+}(q^2)\, , \nonumber\\
F_{0}(q^2)&=&F_{+}(q^2)+\frac{q^2}{m^2_{B_{q^\prime}}-m_S^2}F_{-}(q^2)\, .
\end{eqnarray}

 We carry out the operator product
expansion for the correlation functions $\Pi^i_{\mu}(p,q)$  in the large space-like
momentum region $(p+q)-m_b^2\ll 0$ and the large recoil region of the decaying $B_{q^\prime}$-meson,
which correspond to the small light-cone distance $x^2\approx0$ and are required by the validity of the operator product
expansion. We contract the $b$-quarks in the correlation functions $\Pi^i_{\mu}(p,q)$ with Wick's theorem,
then substantiate  the
free $b$-quark propagator and the corresponding $S$-meson LCDAs into the correlation functions
$\Pi^i_{\mu}(p,q)$   and complete the integrals both in the coordinate space and momentum space
to obtain
\begin{eqnarray}
\Pi^A_{\mu}(p,q)&=& im_b^2 \int_0^1 du   \left\{ \frac{\phi(u)}{m_b^2-(q+up)^2}p_\mu-\frac{m_S\phi_s(u)}{m_b\left[m_b^2-(q+up)^2\right]}(q+up)_\mu\right.\nonumber\\
&&-\frac{m_S\phi_\sigma(u)}{3m_b}\left[\frac{1}{m_b^2-(q+up)^2}+\frac{m_b^2}{\left(m_b^2-(q+up)^2\right)^2} \right]p_\mu\nonumber\\
&&+\frac{m_S\phi_\sigma(u)}{6m_b}\left. \left[p_\mu+(q+up)_\mu \frac{d}{du}\right] \frac{1}{m_b^2-(q+up)^2}\right\}+\cdots \, , \\
\Pi^B_{\mu}(p,q)&=& m_b \int_0^1 du   \left\{ \frac{\phi(u)}{m_b^2-(q+up)^2}-\frac{m_Sm_b\phi_\sigma(u)}{3\left[m_b^2-(q+up)^2\right]^2}\right\}\left(q^2p_\mu- q\cdot p q_\mu \right)+\cdots \, .
\end{eqnarray}

Now we take a short digression to  discuss the
LCDAs of the related two-quark scalar mesons. Let us write down the definitions for
 the twist-2 and twist-3 LCDAs $\phi(u,\mu)$, $\phi_s(u,\mu)$ and $\phi_\sigma(u,\mu)$,
\begin{eqnarray}
\langle S(p)|\bar{q}(x)\gamma_\mu q^\prime(y)|0 \rangle &=& p_\mu
\int^1_0 due^{iup\cdot x+\bar{u}p\cdot y}\phi(u,\mu)\, ,\nonumber\\
\langle S(p)|\bar{q}(x)q^\prime(y)|0\rangle &=& m_S \int_0^1 du \, e^{i(u p\cdot x+ \bar{u}p \cdot y)} \phi_{s}(u,\mu)  \, , \nonumber\\
\langle S(p)|\bar{q}(x) \sigma_{\mu \nu} q^\prime(y)|0\rangle &=& -m_{S}(p_{\mu}z_{\nu} -p_{\nu}z_{\mu})  \int_0^1 du \, e^{i(u p\cdot x+ \bar{u}p \cdot y)} \frac{\phi_{\sigma}(u,\mu)}{6} \, ,
\end{eqnarray}
where the $u$ is the fraction of the light-cone momentum of the scalar
meson carried by the $q$-quark,  $\bar{u}=1-u$ and $z=x-y$.
The LCDAs $\phi(u,\mu)$, $\phi_s(u,\mu)$ and $\phi_\sigma(u,\mu)$ can be expanded into a series of Gegenbauer
polynomials $C^{1/2(3/2)}_{m}(u-\bar{u})$ with increasing conformal spin according to the conformal symmetry of the QCD,
\begin{eqnarray}
\phi(u,\mu)&=&\bar{f}_S(\mu)6u\bar{u}\left[B_0(\mu)+\sum_{m=1}^{\infty}B_m(\mu)C^{3/2}_m
(2u-1)\right] \, , \nonumber  \\
\phi_s(u,\mu)&=&\bar{f}_S(\mu)\left[1+\sum_{m=1}^{\infty}B^s_m(\mu)C^{1/2}_m
(2u-1)\right] \, , \nonumber  \\
\phi_\sigma(u,\mu)&=&\bar{f}_S(\mu)6u\bar{u}\left[1+\sum_{m=1}^{\infty}B^\sigma_m(\mu)C^{3/2}_m
(2u-1)\right] \, ,
\end{eqnarray}
where the $B_m(\mu)$, $B^s_m(\mu)$ and $B_m^\sigma(\mu)$ are the Gegenbauer moments \cite{LCSR-CFS,CFS,YMWang}.
The twist-2 LCDA is antisymmetric under the interchange
$u\leftrightarrow \bar{u}$ in the flavor $SU(3)$ symmetry limit, and the zeroth Gegenbauer moment
$B_0$ vanishes in the flavor $SU(3)$ symmetry limit,  the odd
Gegenbauer moments are dominant. In this article, we  take into account the first two odd
moments $B_1$ and $B_3$ and set $B^s_m=B^\sigma_m=0$ \cite{ChengHY}.

\begin{figure}
 \centering
 \includegraphics[totalheight=7cm,width=15cm]{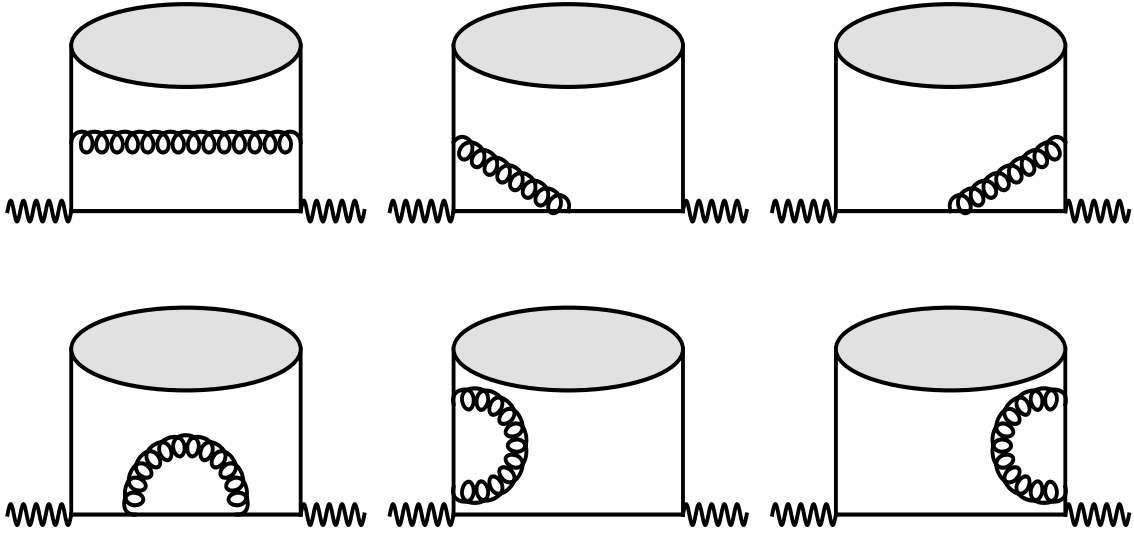}
    \caption{The six Feynman diagrams contribute to the perturbative ${\mathcal{O}}(\alpha_s)$ corrections, they are denoted by $a$, $b$, $c$, $d$, $e$, $f$ sequentially in the text.}
\end{figure}

Then we calculate the perturbative ${\mathcal{O}}(\alpha_s)$ contributions to
  the twist-2 terms, while the perturbative ${\mathcal{O}}(\alpha_s)$
 corrections to the twist-3 terms are beyond the present work, as the contributions of the twist-3 terms are suppressed by the factor $m_S/m_b$ (or $m_bm_S/M^2$ after the Borel transformation) and play a less important role, see Eqs.(7-8).
The  six  Feynman diagrams which determine the perturbative ${\mathcal{O}}(\alpha_s)$ corrections
are shown explicitly in Fig.1.  For simplicity, we perform the calculations
in  the Feynman gauge,  introduce convenient dimensionless
variables $r_1 = q^2/m_b^2$,
$r_2=(p+q)^2/m_b^2$ (or $s/m_b^2$), $\rho=r_1+u(r_2-r_1)$, and take the approximation $m_S^2/m_b^2\approx0$. In calculations, we regularize both the ultraviolet
  and collinear (or infrared) divergences with
dimensional regularization by setting $D=4-2\varepsilon_{\rm UV}=4+2\varepsilon_{\rm IR}$, and perform the renormaliztion
in the $\overline{MS}$ scheme
with  totally anti-commuting $\gamma_5$. In the following, we write down the contributions of the diagrams  $a$, $b$, $c$, $d$, $e$ and $f$,
respectively,
\begin{eqnarray}
\Pi_\mu^{A,a}(p,q)&=&-\frac{C_Fg_s^2m_b}{4} \int \frac{d^Dk}{(2\pi)^D}\int_0^1 du\frac{\phi(u)}{k^2}\nonumber\\
&&\left\{ \not\!p \gamma^\lambda \frac{{\not\!k}+u{\not\!p}}{(k+up)^2} \gamma_\mu\gamma_5\frac{{\not\!k}+\not\!q+u{\not\!p}+m_b}{(k+q+up)^2-m_b^2} \gamma_5 \frac{{\not\!k}-\bar{u}{\not\!p}}{(k-\bar{u}p)^2}\gamma_\lambda\right\}\, ,\nonumber\\
&=&-\frac{iC_F\alpha_s }{2\pi} \int_0^1 du \phi(u) \left\{ \left[ \frac{\log(1-r_2)}{r_2}-\frac{\log(1-r_1)}{r_1}\right]q_\mu \right.\nonumber\\
&&+\left[\frac{u}{1-r_1}+\frac{\bar{u}}{1-r_2}+\frac{\log(1-r_2)-\log(1-r_1)}{r_2-r_1} \right]\left(\frac{1}{\hat{\varepsilon}_{\rm IR}}+\log\frac{m_b^2}{\mu^2}\right)p_\mu \nonumber\\
&&\left.+\left[\frac{G(r_2)-G(r_1)}{r_2-r_1}+\frac{\log(1-r_2)}{r_2}+u\frac{2\log(1-r_1)+1}{1-r_1}+\bar{u}\frac{2\log(1-r_2)+1}{1-r_2}\right]p_\mu\right\} \, , \nonumber\\
\end{eqnarray}
\begin{eqnarray}
\Pi_\mu^{A,b+c}(p,q)&=&-\frac{C_Fg_s^2m_b}{4} \int \frac{d^Dk}{(2\pi)^D}\int_0^1 du\frac{\phi(u)}{k^2}\nonumber\\
&&\left\{ \not\!p \gamma^\lambda \frac{{\not\!k}+u{\not\!p}}{(k+up)^2} \gamma_\mu\gamma_5\frac{{\not\!k}+\not\!q+u{\not\!p}+m_b}{(k+q+up)^2-m_b^2}\gamma_\lambda\frac{{\not\!q}+u{\not\!p}+m_b}{(q+up)^2-m_b^2} \gamma_5 \right.\nonumber\\
&&\left.+ {\not\!p} \gamma_\mu\gamma_5  \frac{{\not\!q}+u{\not\!p}+m_b}{(q+up)^2-m_b^2}\gamma^\lambda \frac{{\not\!k}+\not\!q+u{\not\!p}+m_b}{(k+q+up)^2-m_b^2}\gamma_5\frac{{\not\!k}-\bar{u}{\not\!p}}{(k-\bar{u}p)^2} \gamma_\lambda \right\}\, ,\nonumber\\
&=&\frac{iC_F\alpha_s }{2\pi} \int_0^1 du \frac{\phi(u)}{1-\rho} \left\{-\frac{1}{2u} \left[ \frac{(1-\rho)F(\rho)}{\rho^2}-\frac{(1-r_1)F(r_1)}{r_1^2}\right]q_\mu \right.\nonumber\\
&&+\left[\frac{F(\rho)-F(r_1)}{u(r_2-r_1)}-\frac{F(\rho)-F(r_2)}{\bar{u}(r_2-r_1)}+2\log(1-\rho)\right]\left(\frac{1}{\hat{\varepsilon}_{\rm IR}}+\log\frac{m_b^2}{\mu^2}\right)p_\mu \nonumber\\
&&+\left[\frac{5}{2}\left(\frac{1}{\hat{\varepsilon}_{\rm UV}}-\log\frac{m_b^2}{\mu^2}\right)-\frac{(1-\rho)F(\rho)}{2\rho^2}+2\log^2(1-\rho)+2{\rm Li_2(\rho)}\right.\nonumber\\
&&\left.+3+\frac{F(\rho)}{\rho} +\frac{F(r_2)}{r_2}\right]p_\mu\nonumber\\
&&  +\left[\frac{(1-\rho)G(\rho)-(1-r_1)G(r_1)-2F(\rho)+2F(r_1)}{u(r_2-r_1)}\right.\nonumber\\
&&\left.\left.-\frac{(1-\rho)G(\rho)-(1-r_2)G(r_2)-2F(\rho)+2F(r_2)}{\bar{u}(r_2-r_1)}\right]p_\mu\right\} \, ,
\end{eqnarray}
\begin{eqnarray}
\Pi_\mu^{A,d}(p,q)&=&-\frac{C_Fg_s^2m_b}{4} \int \frac{d^Dk}{(2\pi)^D}\int_0^1 du\frac{\phi(u)}{k^2}\nonumber\\
&&\left\{ \not\!p \gamma_\mu\gamma_5 \frac{{\not\!q}+u{\not\!p}+m_b}{(q+up)^2-m_b^2}\gamma^\lambda \frac{{\not\!k}+\not\!q+u{\not\!p}+m_b}{(k+q+up)^2-m_b^2}\gamma_\lambda \frac{{\not\!q}+u{\not\!p}+m_b}{(q+up)^2-m_b^2}\gamma_5 \right\}\, ,\nonumber\\
&=&-\frac{iC_F\alpha_s }{2\pi} \int_0^1 du \frac{\phi(u)\rho}{(1-\rho)^2} \left\{ \frac{1}{\hat{\varepsilon}_{\rm UV}}-\log\frac{m_b^2}{\mu^2}+1-\frac{(1-\rho)F(\rho)}{\rho^2} \right\}p_\mu \, ,
\end{eqnarray}
\begin{eqnarray}
\Pi_\mu^{A,e+f}(p,q)&=&-\frac{C_Fg_s^2m_b}{4} \int \frac{d^Dk}{(2\pi)^D}\int_0^1 du\frac{\phi(u)}{k^2}\nonumber\\
&&\left\{ {\not\!p} \gamma^\lambda \frac{{\not\!k}+u{\not\!p}}{(k+up)^2} \gamma_\lambda\frac{u{\not\!p}}{(up)^2}\gamma_\mu\gamma_5\frac{\not\!q+u{\not\!p}+m_b}{(q+up)^2-m_b^2} \gamma_5 \right.\nonumber\\
&&\left.+{\not\!p} \gamma_\mu\gamma_5\frac{\not\!q+u{\not\!p}+m_b}{(q+up)^2-m_b^2} \gamma_5 \frac{-\bar{u}{\not\!p}}{(-\bar{u}p)^2} \gamma^\lambda\frac{{\not\!k}-\bar{u}{\not\!p}}{(k-\bar{u}p)^2}\gamma_\lambda\right\}\, ,\nonumber\\
&\propto&\frac{1}{\varepsilon_{\rm UV}}-\frac{1}{\varepsilon_{\rm IR}} \, ,
\end{eqnarray}

\begin{eqnarray}
\Pi_\mu^{B,a}(p,q)&=&-\frac{C_Fg_s^2m_b}{4} \int \frac{d^Dk}{(2\pi)^D}\int_0^1 du\frac{\phi(u)}{k^2}\nonumber\\
&&\left\{ \not\!p \gamma^\lambda \frac{{\not\!k}+u{\not\!p}}{(k+up)^2} \sigma_{\mu\nu}\gamma_5q^\nu\frac{{\not\!k}+\not\!q+u{\not\!p}+m_b}{(k+q+up)^2-m_b^2} \gamma_5 \frac{{\not\!k}-\bar{u}{\not\!p}}{(k-\bar{u}p)^2}\gamma_\lambda\right\}\, ,\nonumber\\
&=&-\frac{C_F\alpha_s }{2\pi m_b} \int_0^1 du \frac{\phi(u)}{1-\rho}\left\{\left[\frac{u}{1-r_1}+\frac{\bar{u}}{1-r_2}+\frac{\log(1-r_2)-\log(1-r_1)}{r_2-r_1} \right] \right. \nonumber\\
&&\left(\frac{1}{\hat{\varepsilon}_{\rm IR}}+\log\frac{m_b^2}{\mu^2}\right)+\frac{G(r_2)-G(r_1)}{r_2-r_1}+\frac{1}{r_2-r_1}\left[\frac{F(r_2)}{r_2}-\frac{F(r_1)}{r_1}\right]
\nonumber\\
&&\left.+\frac{u\log(1-r_1)}{r_1}+\frac{\bar{u}\log(1-r_2)}{r_2} +u\frac{2\log(1-r_1)+1}{1-r_1}+\bar{u}\frac{2\log(1-r_2)+1}{1-r_2} \right\} \nonumber\\
&&\left(q^2p_\mu- q\cdot p q_\mu \right)\, ,
\end{eqnarray}
\begin{eqnarray}
\Pi_\mu^{B,b+c}(p,q)&=&-\frac{C_Fg_s^2m_b}{4} \int \frac{d^Dk}{(2\pi)^D}\int_0^1 du\frac{\phi(u)}{k^2}\nonumber\\
&&\left\{ \not\!p \gamma^\lambda \frac{{\not\!k}+u{\not\!p}}{(k+up)^2} \sigma_{\mu\nu}\gamma_5q^\nu\frac{{\not\!k}+\not\!q+u{\not\!p}+m_b}{(k+q+up)^2-m_b^2}\gamma_\lambda\frac{{\not\!q}+u{\not\!p}+m_b}{(q+up)^2-m_b^2} \gamma_5 \right.\nonumber\\
&&\left.+ {\not\!p} \sigma_{\mu\nu}\gamma_5q^\nu  \frac{{\not\!q}+u{\not\!p}+m_b}{(q+up)^2-m_b^2}\gamma^\lambda \frac{{\not\!k}+\not\!q+u{\not\!p}+m_b}{(k+q+up)^2-m_b^2}\gamma_5\frac{{\not\!k}-\bar{u}{\not\!p}}{(k-\bar{u}p)^2} \gamma_\lambda \right\}\, ,\nonumber\\
&=&\frac{C_F\alpha_s }{2\pi m_b} \int_0^1 du \frac{\phi(u)}{1-\rho} \left\{ \left[\frac{F(\rho)-F(r_1)}{u(r_2-r_1)}-\frac{F(\rho)-F(r_2)}{\bar{u}(r_2-r_1)}+2\log(1-\rho)\right]\right.
 \nonumber\\
&&\left(\frac{1}{\hat{\varepsilon}_{\rm IR}}+\log\frac{m_b^2}{\mu^2}\right)+\left[2\left(\frac{1}{\hat{\varepsilon}_{\rm UV}}-\log\frac{m_b^2}{\mu^2}\right)+2\log^2(1-\rho)+2{\rm Li_2(\rho)}+3\right.
\nonumber\\
&&\left.+\frac{2F(\rho)}{\rho} \right]+\frac{1}{u(r_2-r_1)}\left[ \frac{F(\rho)}{\rho}-\frac{F(r_1)}{r_1} \right] +\frac{1}{\bar{u}(r_2-r_1)}\left[ \frac{F(\rho)}{\rho}-\frac{F(r_2)}{r_2} \right] \nonumber\\
&& +\left[\frac{(1-\rho)G(\rho)-(1-r_1)G(r_1)-3F(\rho)+3F(r_1)}{u(r_2-r_1)}\right.\nonumber\\
&&\left.\left.-\frac{(1-\rho)G(\rho)-(1-r_2)G(r_2)-F(\rho)+F(r_2)}{\bar{u}(r_2-r_1)}\right]\right\}  \left(q^2p_\mu- q\cdot p q_\mu \right)\, ,
\end{eqnarray}
\begin{eqnarray}
\Pi_\mu^{B,d}(p,q)&=&-\frac{C_Fg_s^2m_b}{4} \int \frac{d^Dk}{(2\pi)^D}\int_0^1 du\frac{\phi(u)}{k^2}\nonumber\\
&&\left\{ \not\!p \sigma_{\mu\nu}\gamma_5q^\nu \frac{{\not\!q}+u{\not\!p}+m_b}{(q+up)^2-m_b^2}\gamma^\lambda \frac{{\not\!k}+\not\!q+u{\not\!p}+m_b}{(k+q+up)^2-m_b^2}\gamma_\lambda \frac{{\not\!q}+u{\not\!p}+m_b}{(q+up)^2-m_b^2}\gamma_5 \right\}\, ,\nonumber\\
&=&-\frac{C_F\alpha_s }{4\pi m_b} \int_0^1 du \frac{\phi(u)}{(1-\rho)^2} \left\{ (7-\rho)\left[\frac{1}{\hat{\varepsilon}_{\rm UV}}-\log\frac{m_b^2}{\mu^2}+\frac{F(\rho)}{\rho}\right]\right.\nonumber\\
&&\left.+4-\frac{(1+\rho)F(\rho)}{\rho^2} \right\}\left(q^2p_\mu- q\cdot p q_\mu \right)\, ,
\end{eqnarray}
\begin{eqnarray}
\Pi_\mu^{B,e+f}(p,q)&=&-\frac{C_Fg_s^2m_b}{4} \int \frac{d^Dk}{(2\pi)^D}\int_0^1 du\frac{\phi(u)}{k^2}\nonumber\\
&&\left\{ {\not\!p} \gamma^\lambda \frac{{\not\!k}+u{\not\!p}}{(k+up)^2} \gamma_\lambda\frac{u{\not\!p}}{(up)^2}\sigma_{\mu\nu}\gamma_5q^\nu\frac{\not\!q+u{\not\!p}+m_b}{(q+up)^2-m_b^2} \gamma_5 \right.\nonumber\\
&&\left.+{\not\!p} \sigma_{\mu\nu}\gamma_5q^\nu\frac{\not\!q+u{\not\!p}+m_b}{(q+up)^2-m_b^2} \gamma_5 \frac{-\bar{u}{\not\!p}}{(-\bar{u}p)^2} \gamma^\lambda\frac{{\not\!k}-\bar{u}{\not\!p}}{(k-\bar{u}p)^2}\gamma_\lambda\right\}\, ,\nonumber\\
&\propto&\frac{1}{\varepsilon_{\rm UV}}-\frac{1}{\varepsilon_{\rm IR}} \, ,
\end{eqnarray}

where
\begin{eqnarray}
\frac{1}{\hat{\varepsilon}_{\rm UV}}&=&\frac{1}{\varepsilon_{\rm UV}}-\gamma_E+\log4\pi\, , \nonumber\\
\frac{1}{\hat{\varepsilon}_{\rm IR}}&=&\frac{1}{\varepsilon_{\rm IR}}+\gamma_E-\log4\pi\, , \nonumber\\
F(\rho)&=&(1-\rho)\log(1-\rho)+\rho\, ,\nonumber\\
G(\rho)&=&\log^2(1-\rho)+{\rm Li_2}(\rho)+\log(1-\rho)\, ,
\end{eqnarray}
${\rm Li_2}(x)=-\int_0^x dt \frac{\log(1-t)}{t}$ and $C_F=\frac{4}{3}$. The terms proportional to the  ultraviolet divergence $\frac{1}{\hat{\varepsilon}_{\rm UV}}$ are eliminated through renormalization,
while the terms  proportional to the  infrared (or collinear) divergence $\frac{1}{\hat{\varepsilon}_{\rm IR}}+\log\frac{m_b^2}{\mu^2}$ are absorbed in the
twist-2 LCDA $\phi(u,\mu)$. The terms proportional to $\frac{1}{\bar{u}}$ are potentially divergent at the end point $u=1$ after performing the quark-hadron duality, we absorb those terms into the  twist-2 LCDA $\phi(u,\mu)$. We introduce the notations $\widetilde{\Pi}_\mu^{A,\alpha_s}(p,q)$, $\widetilde{\Pi}_\mu^{B,\alpha_s}(p,q)$, $\widetilde{\Pi}_\mu^{A,a,b,c,d}(p,q)$, $\widetilde{\Pi}_\mu^{B,a,b,c,d}(p,q)$, $\widetilde{\Pi}_{+}(q^2,(p+q)^2)$, $\widetilde{\Pi}_{+-}(q^2,(p+q)^2)$, $\widetilde{\Pi}_{T}(q^2,(p+q)^2)$  to denote the renormalized  correlation functions,
\begin{eqnarray}
\widetilde{\Pi}_\mu^{A,\alpha_s}(p,q)&=&\widetilde{\Pi}_\mu^{A,a}(p,q)+\widetilde{\Pi}_\mu^{A,b+c}(p,q)+\widetilde{\Pi}_\mu^{A,d}(p,q)\, ,\nonumber\\
&=&\widetilde{\Pi}_{+}(q^2,(p+q)^2)p_\mu+\widetilde{\Pi}_{+-}(q^2,(p+q)^2)q_\mu \, ,\nonumber\\
\widetilde{\Pi}_\mu^{B,\alpha_s}(p,q)&=&\widetilde{\Pi}_\mu^{B,a}(p,q)+\widetilde{\Pi}_\mu^{B,b+c}(p,q)+\widetilde{\Pi}_\mu^{B,d}(p,q)\, ,\nonumber\\
&=&\widetilde{\Pi}_{T}(q^2,(p+q)^2)\left(q^2p_\mu-q\cdot p q_\mu \right)\, .
\end{eqnarray}
Then we obtain the QCD  spectral densities through the dispersion relation,
\begin{eqnarray}
\widetilde{\Pi}_{i}(q^2,(p+q)^2)&=&\int_{m_b^2}^{s_0}ds \frac{\rho_{i}^{\alpha_s}(s)}{s-(p+q)^2}+\int_{s_0}^{\infty}ds \frac{\rho_{i}^{\alpha_s}(s)}{s-(p+q)^2}\, ,
\end{eqnarray}
where $i=+,+-,T$ and
\begin{eqnarray}
\rho_{+}^{\alpha_s}(s)&=&\frac{{\rm Im }\widetilde{\Pi}_{+}(q^2,s)}{\pi} \, , \nonumber\\
&=&\frac{C_F\alpha_s}{2\pi}\int_{\bar{\Delta}}^1 du \phi(u)\left\{ \frac{1}{r_2}+\frac{2\bar{u}}{1-r_2}-\frac{\log r_2-2\log(r_2-1)-1}{r_2-r_1}\right.\nonumber\\
&&+\Theta(\rho-1)\left[\frac{1-\rho}{2\rho^2}-\frac{4\log(\rho-1)}{1-\rho}\mid_{+} +\frac{2\log\rho}{1-\rho}\mid_{+}-\frac{2}{\rho}-\frac{1-r_2}{r_2}\frac{1}{1-\rho}\mid_{+} \right. \nonumber\\
&&\left.+\frac{\log\rho+1-2\log(\rho-1)}{u(r_2-r_1)} \right]\nonumber\\
&&+\Theta(1-\rho)\left[-\frac{1-r_2}{r_2}\frac{1}{1-\rho}\mid_{+} \right]\nonumber\\
&&+\delta(1-\rho) \left[-\frac{5}{2}\log\frac{m_b^2}{\mu^2} +2\log^2(r_2-1)+2{\rm Li_2}(1-r_2) -\frac{\pi^2}{3}+6\right.\nonumber\\
&&+\frac{1-r_2}{r_2}\left[ 2\log(r_2-1)-\log(1-r_1)\right]+\left(1-\log\frac{m_b^2}{\mu^2} \right)\left(1+\frac{1}{r_2-r_1}\frac{\overleftarrow{d}}{du}\right) \nonumber\\
&&\left.\left.+\frac{2F(r_1)-(1-r_1)G(r_1)-2}{u(r_2-r_1)}\right]\right\}\, ,
\end{eqnarray}
\begin{eqnarray}
\rho_{+-}^{\alpha_s}(s)&=&\frac{{\rm Im }\widetilde{\Pi}_{+-}(q^2,s)}{\pi} \, , \nonumber\\
&=&\frac{C_F\alpha_s}{2\pi}\int_{\bar{\Delta}}^1 du \phi(u)\left\{ \frac{1}{r_2}-\Theta(\rho-1) \frac{\rho-1}{2u\rho^2} +\delta(1-\rho)\frac{(1-r_1)F(r_1)}{2ur_1^2}\right\}\, ,
\end{eqnarray}
\begin{eqnarray}
\rho_{T}^{\alpha_s}(s)&=&\frac{{\rm Im }\widetilde{\Pi}_{T}(q^2,s)}{\pi} \, , \nonumber\\
&=&\frac{C_F\alpha_s}{2\pi m_b}\int_{\bar{\Delta}}^1 du \phi(u)\left\{ \frac{2\log(r_2-1)-\log r_2+1}{r_2-r_1} -\frac{r_2-1}{r_2(r_2-r_1)}+\frac{\bar{u}}{r_2}+\frac{2\bar{u}}{1-r_2}\right. \nonumber\\
&&+\Theta(\rho-1)\left[-\frac{3}{2\rho}-\frac{4\log(\rho-1)}{1-\rho}\mid_{+} +\frac{2\log\rho}{1-\rho}\mid_{+} +\frac{\log\rho-2\log(\rho-1)-1}{u(r_2-r_1)}\right. \nonumber\\
&&\left.+\left(3-\frac{1}{\rho} \right)\frac{1}{u(r_2-r_1)}+\frac{5\rho-1}{2\rho^2}\frac{1}{1-\rho}\mid_{+} \right]+\delta(1-\rho)\left[ \frac{9}{2}-\frac{\pi^2}{3}+2\log^2(r_2-1)\right. \nonumber\\
&&+2{\rm Li_2}(1-r_2)-\frac{3}{2}\log\frac{m_b^2}{\mu^2}+\left(3F(r_1)-(1-r_1)G(r_1)-2-\frac{F(r_1)}{r_1} \right)\frac{1}{u(r_2-r_1)} \nonumber\\
&&\left.\left.-2\log(r_2-1)-3\log\frac{m_b^2}{\mu^2}\frac{1}{r_2-r_1}\overleftarrow{\frac{d}{du}}+\frac{9}{2}\frac{1}{r_2-r_1}\overleftarrow{\frac{d}{du}}-
\frac{1}{2\rho}\frac{1}{r_2-r_1}\overleftarrow{\frac{d}{du}}
\right]\right\}\, ,
\end{eqnarray}
$\bar{\Delta}=\frac{m_b^2-q^2}{s_0-q^2}$, $\phi(\rho)\frac{f(\rho)}{1-\rho}\mid_{+}=\frac{f(\rho)}{1-\rho}\left[\phi(\rho)-\phi(1)\right]$, and $\Theta(x)=1$ for $x\geq 0$, the $\overleftarrow{\frac{d}{du}}$ does not act on the $\delta(1-\rho)$.

We take quark-hadron duality below the continuum thresholds $s_0$, and perform the  Borel transformation   with respect to the variable
$P^{\prime2}=-(p+q)^2$ to obtain the LCSR,
\begin{eqnarray}
F_{+}(q^2)&=&-\frac{m_b^2}{2f_{B_{q^\prime}}m_{B_{q^\prime}}^2}\exp\left( \frac{m_{B_{q^\prime}}^2}{M^2}\right) \int_{\Delta}^1du\left\{\frac{\phi(u)}{u}-\frac{m_S\phi_s(u)}{m_b} -\frac{m_S\phi_\sigma(u)}{3um_b}\left(1+\frac{m_b^2}{u M^2} \right)\right. \nonumber\\
&&\left.+\frac{m_S\phi_\sigma(u)}{6um_b}\left( 1+u^2\frac{d}{du} \frac{1}{u}\right) \right\}\exp\left(-\frac{m_b^2+u\bar{u}m_S^2-\bar{u}q^2}{uM^2} \right)\nonumber\\
&&-\frac{m_b^2}{2f_{B_{q^\prime}}m_{B_{q^\prime}}^2}\exp\left( \frac{m_{B_{q^\prime}}^2}{M^2}\right)  \int_{m_b^2}^{s_0}ds\rho^{\alpha_s}_{+}(s)\exp\left( -\frac{s}{M^2}\right) \, ,
\end{eqnarray}
\begin{eqnarray}
F_{+}(q^2)+F_{-}(q^2)&=&-\frac{m_b^2}{f_{B_{q^\prime}}m_{B_{q^\prime}}^2}\exp\left( \frac{m_{B_{q^\prime}}^2}{M^2}\right) \frac{m_S}{m_b} \int_{\Delta}^1du\left\{-\frac{\phi_s(u)}{u} +\frac{\phi_\sigma(u)}{6}\frac{d}{du} \frac{1}{u}\right\}\nonumber\\
&&\exp\left(-\frac{m_b^2+u\bar{u}m_S^2-\bar{u}q^2}{uM^2} \right) \nonumber\\
&&-\frac{m_b^2}{f_{B_{q^\prime}}m_{B_{q^\prime}}^2}\exp\left( \frac{m_{B_{q^\prime}}^2}{M^2}\right) \int_{m_b^2}^{s_0}ds\rho^{\alpha_s}_{+-}(s)\exp\left( -\frac{s}{M^2}\right) \, ,
\end{eqnarray}
\begin{eqnarray}
F_{T}(q^2)&=&-\frac{m_{B_{q^\prime}}+m_S}{2f_{B_{q^\prime}}m_{B_{q^\prime}}^2}\exp\left( \frac{m_{B_{q^\prime}}^2}{M^2}\right) \int_{\Delta}^1du\left\{\frac{ m_b\phi(u)}{u} -\frac{ m_b^2m_S\phi_\sigma(u)}{3u^2M^2} \right\}\nonumber\\
&&\exp\left(-\frac{m_b^2+u\bar{u}m_S^2-\bar{u}q^2}{uM^2} \right) \nonumber\\
&&-\frac{m_{B_{q^\prime}}+m_S}{2f_{B_{q^\prime}}m_{B_{q^\prime}}^2}\exp\left( \frac{m_{B_{q^\prime}}^2}{M^2}\right) \int_{m_b^2}^{s_0}ds\rho^{\alpha_s}_{T}(s)\exp\left( -\frac{s}{M^2}\right) \, ,
\end{eqnarray}
where $\Delta=\left[ \sqrt{(s-q^2-m_S^2)^2+4m_S^2(m_b^2-q^2)}-(s-q^2-m_S^2)\right]/2m_S^2$, numerically $\Delta\approx \bar{\Delta}$.

\section{Numerical results and discussions}
The hadronic masses are taken as
$m_{B^\pm}=5279.25\,\rm{MeV}$,
$m_{B^0}=5279.55\,\rm{MeV}$,
$m_{B_s}=5366.7\,\rm{MeV}$,
$m_{a_0(980)}=980\,\rm{MeV}$,
$m_{\kappa(800)}=682\,\rm{MeV}$,
$m_{f_0(980)}=990\,\rm{MeV}$,
$m_{a_0(1450)}=1474\,\rm{MeV}$,
$m_{K_0^*(1430)}=1425\,\rm{MeV}$,
$m_{f_0(1500)}=1505\,\rm{MeV}$
 from the Particle Data Group
\cite{PDG}.
The threshold parameters and the decay constants of the $B_{q^\prime}$ mesons are taken from the two-point QCD sum rules $s^0_{B}=(33\pm1)\,\rm{GeV}^2$, $s^0_{B_s}=(35\pm1)\,\rm{GeV}^2$, $f_B=(190\pm17)\,\rm{MeV}$ and $f_{B_s}=(233\pm17)\,\rm{MeV}$  \cite{WangJHEP}.

In the article, we take the $\overline{MS}$ mass $m_{b}(m_b)=(4.18\pm0.03)\,\rm{GeV}$
 from the Particle Data Group \cite{PDG}, and take into account
the energy-scale dependence of  the $\overline{MS}$ mass from the renormalization group equation,
\begin{eqnarray}
m_b(\mu)&=&m_b(m_b)\left[\frac{\alpha_{s}(\mu)}{\alpha_{s}(m_b)}\right]^{\frac{12}{23}} \, ,\nonumber\\
\alpha_s(\mu)&=&\frac{1}{b_0t}\left[1-\frac{b_1}{b_0^2}\frac{\log t}{t} +\frac{b_1^2(\log^2{t}-\log{t}-1)+b_0b_2}{b_0^4t^2}\right]\, ,
\end{eqnarray}
  where $t=\log \frac{\mu^2}{\Lambda^2}$, $b_0=\frac{33-2n_f}{12\pi}$, $b_1=\frac{153-19n_f}{24\pi^2}$, $b_2=\frac{2857-\frac{5033}{9}n_f+\frac{325}{27}n_f^2}{128\pi^3}$,  $\Lambda=213\,\rm{MeV}$, $296\,\rm{MeV}$  and  $339\,\rm{MeV}$ for the flavors  $n_f=5$, $4$ and $3$, respectively  \cite{PDG}.
 The scale evolution behaviors of the LCDAs $\phi(u,\mu)$, $\phi_s(u,\mu)$ and $\phi_\sigma(u,\mu)$ are also determined by the renormalization group equation,
\begin{eqnarray}
\bar f_S (\mu)&=&\bar f_S(\mu_0)\Bigg(
\frac{\alpha_s(\mu_0)}{\alpha_s(\mu)}
\Bigg)^{\frac{4}{b}},\nonumber\\
B_m(\mu)&=& B_m(\mu_0)
  \left(\frac{\alpha_s(\mu_0)}{\alpha_s(\mu)}\right)^{-\frac{\gamma_{m}+4}{b}},
  \end{eqnarray}
where $b=(33-2n_f)/3$, and the one-loop anomalous dimensions \cite{Gamma} is
  \begin{eqnarray}
  \gamma_{m}= C_F
  \left(1-\frac{2}{(m+1)(m+2)}+4 \sum_{j=2}^{m+1}
  \frac{1}{j}\right)\, .
  \end{eqnarray}
 We take the values of the  parameters in the LCDAs $\phi(u,\mu)$, $\phi_s(u,\mu)$ and $\phi_\sigma(u,\mu)$ as
 $\bar{f}_{a_0(980)}=(365\pm20)\,\rm{MeV}$, $B_1^{a_0(980)}=-0.93 \pm 0.10$, $B_3^{a_0(980)}=0.14 \pm 0.08$,
 $\bar{f}_{\kappa(800)}=(340\pm20)\,\rm{MeV}$, $B_1^{\kappa(800)}=-0.92 \pm 0.11$, $B_3^{\kappa(800)}=0.15\pm 0.09$,
 $\bar{f}_{f_0(980)}=(370\pm20)\,\rm{MeV}$,  $B_1^{f_0(980)}=-0.78 \pm 0.08$, $B_3^{f_0(980)}=0.02 \pm 0.07$,
 $\bar{f}_{a_0(1450)}=(460\pm50)\,\rm{MeV}$, $B_1^{a_0(1450)}=-0.58 \pm 0.12$, $B_3^{a_0(1450)}=-0.49 \pm 0.15$,
 $\bar{f}_{K_0^*(1430)}=(445\pm50)\,\rm{MeV}$, $B_1^{K_0^*(1430)}=-0.57 \pm 0.13$, $B_3^{K_0^*(1430)}=-0.42 \pm 0.22$,
 $\bar{f}_{f_0(1500)}=(490\pm50)\,\rm{MeV}$,  $B_1^{f_0(1500)}=-0.48 \pm 0.11$, $B_3^{f_0(1500)}=-0.37 \pm 0.20$
  from the two-point QCD sum rules
at the energy scale $\mu_0=1\,\rm{GeV}$ \cite{ChengHY}.

We evolve all the scale dependent quantities to the energy scale $\mu=\sqrt{m_B^2-m_b^2}\approx 2.4\,\rm{GeV}$ to calculate the transition form-factors.
The constituent quarks of the scalar mesons
are in essence off-shell or far from their mass-shell by the virtuality of the order $m_S^2$ as carrying the total momentum of the scalar mesons,
 which have a large light-cone momentum, we should take the energy scale $\mu\geq m_S$. The validity of the operator product expansion requires
$0\leq q^2<(m_b-m_S)^2-2(m_b-m_S)\chi$, where the $\chi$ is a typical hadronic scale of roughly 500 MeV.
The available regions are    $0\leq q^2<11 \rm{GeV}^2$ for a scalar
mesons $a_0(980)$, $\kappa(800)$, $f_0(980)$ and $0\leq q^2<8 \rm{GeV}^2$ for a scalar
mesons $a_0(1450)$, $K_0^*(1430)$, $f_0(1500)$.

In Figs.2-3, we plot the central values of the transition form-factors $F_{+}(0)$, $-F_{-}(0)$ and $F_T(0)$ with variations of the Borel parameter $M^2$. From the figures, we can see that the   form-factors $F_{+}(0)$, $-F_{-}(0)$ and $F_T(0)$ for all the transitions  $B-a_0(980)$, $B-\kappa(800)$, $B_s-\kappa(800)$, $B_s-f_0(980)$, $B-a_0(1450)$, $B-K_0^*(1430)$, $B_s-K_0^*(1430)$ and $B_s-f_0(1500)$ are rather  stable with variations of the Borel parameter $M^2$ at the region $M^2=(8-12)\,\rm{GeV}^2$.  So in the article, we take the Borel parameter as $M^2=(8-12)\,\rm{GeV}^2$.

\begin{figure}
 \centering
 \includegraphics[totalheight=4cm,width=4.8cm]{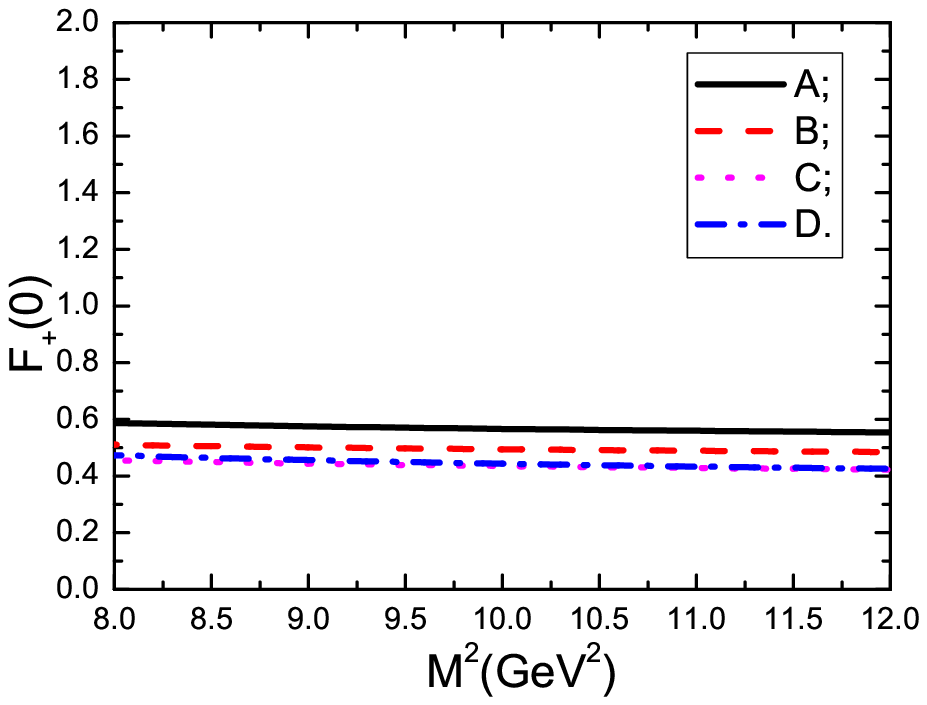}
 \includegraphics[totalheight=4cm,width=4.8cm]{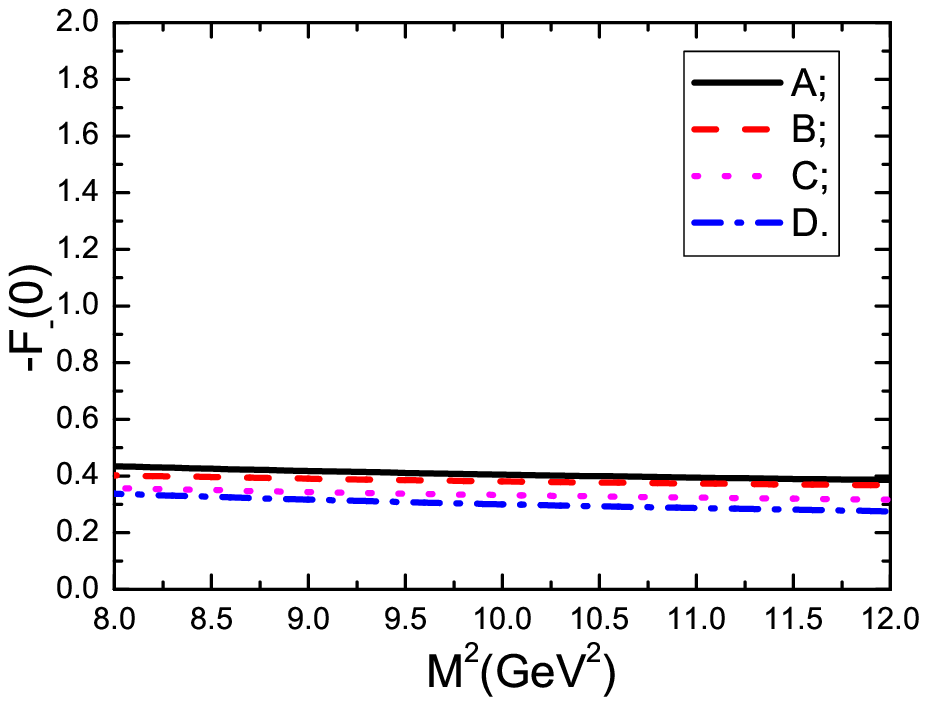}
 \includegraphics[totalheight=4cm,width=4.8cm]{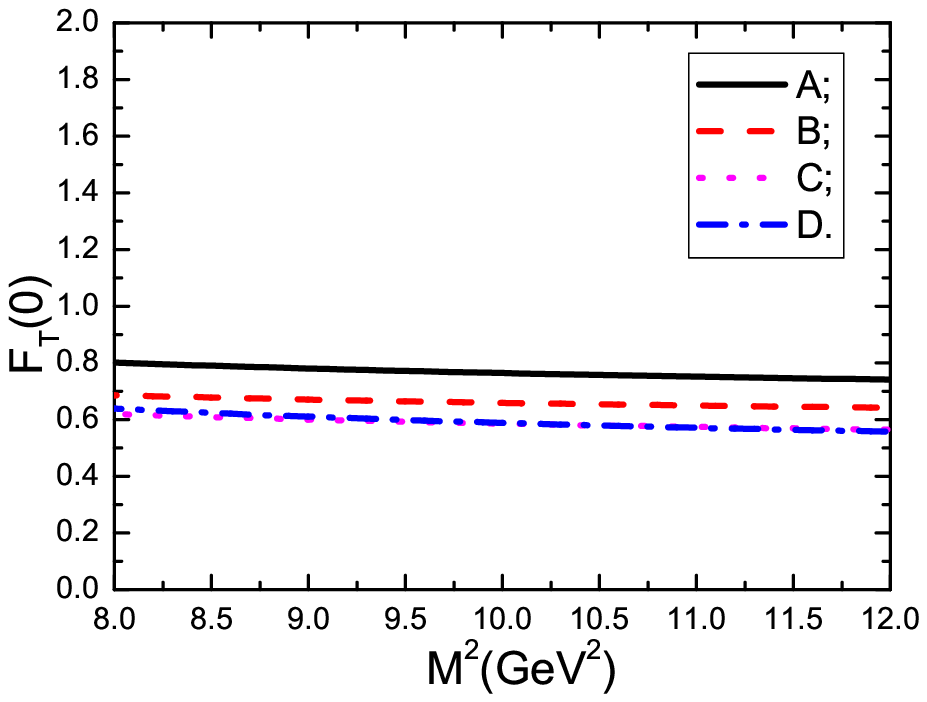}
    \caption{The central values of the form-factors $F_{+}(0)$, $F_{-}(0)$ and $F_T(0)$ with variations of the Borel parameter $M^2$, where $A$, $B$, $C$ and $D$ denote
    the transitions $B-a_0(980)$, $B-\kappa(800)$, $B_s-\kappa(800)$ and $B_s-f_0(980)$, respectively.  }
\end{figure}
\begin{figure}
 \centering
 \includegraphics[totalheight=4cm,width=4.8cm]{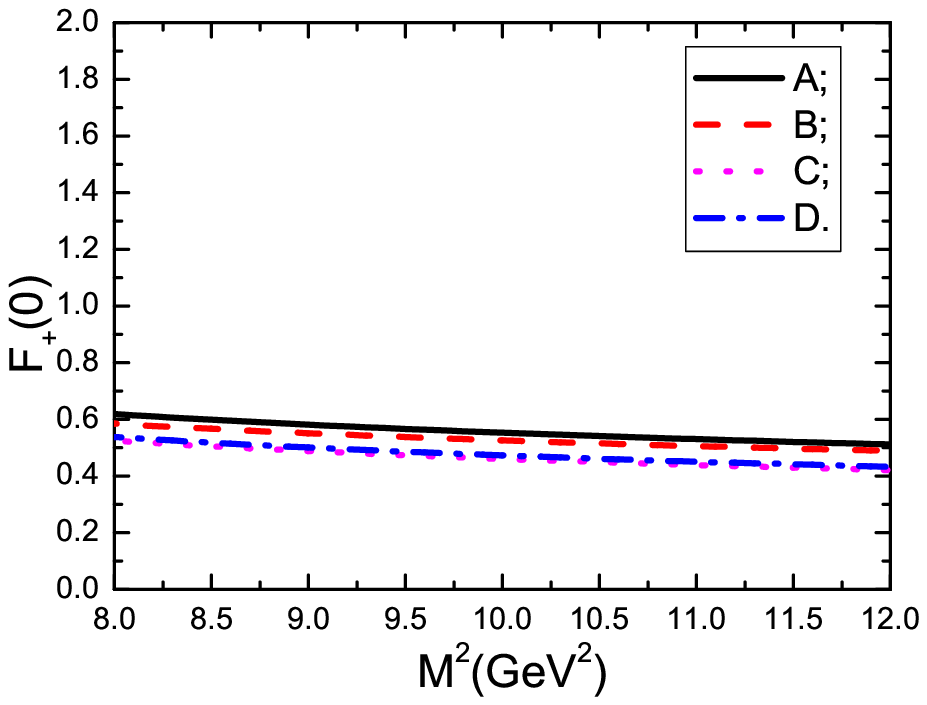}
 \includegraphics[totalheight=4cm,width=4.8cm]{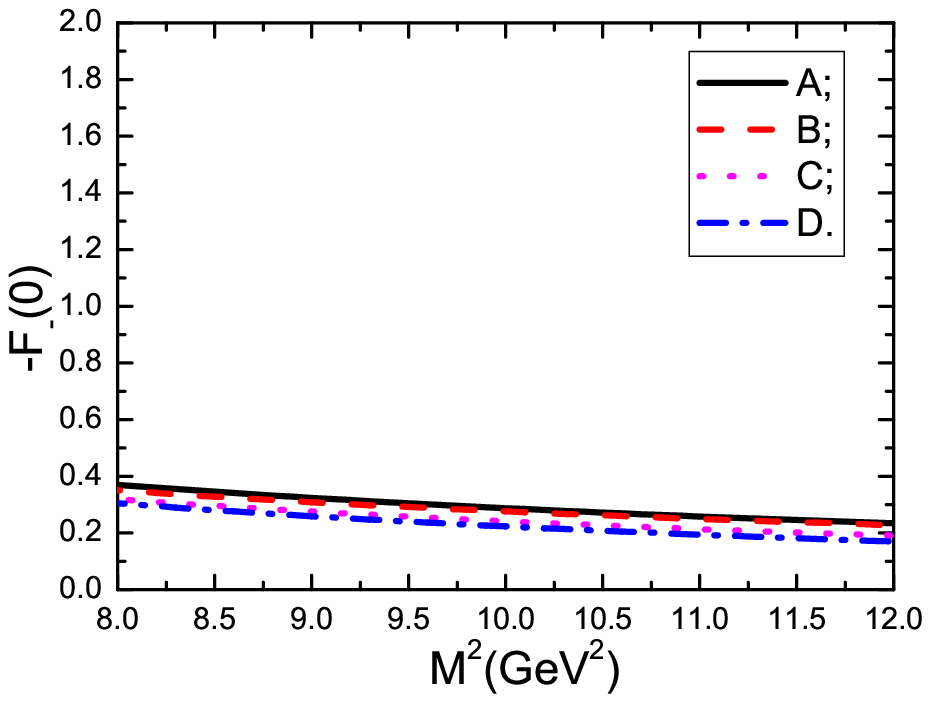}
 \includegraphics[totalheight=4cm,width=4.8cm]{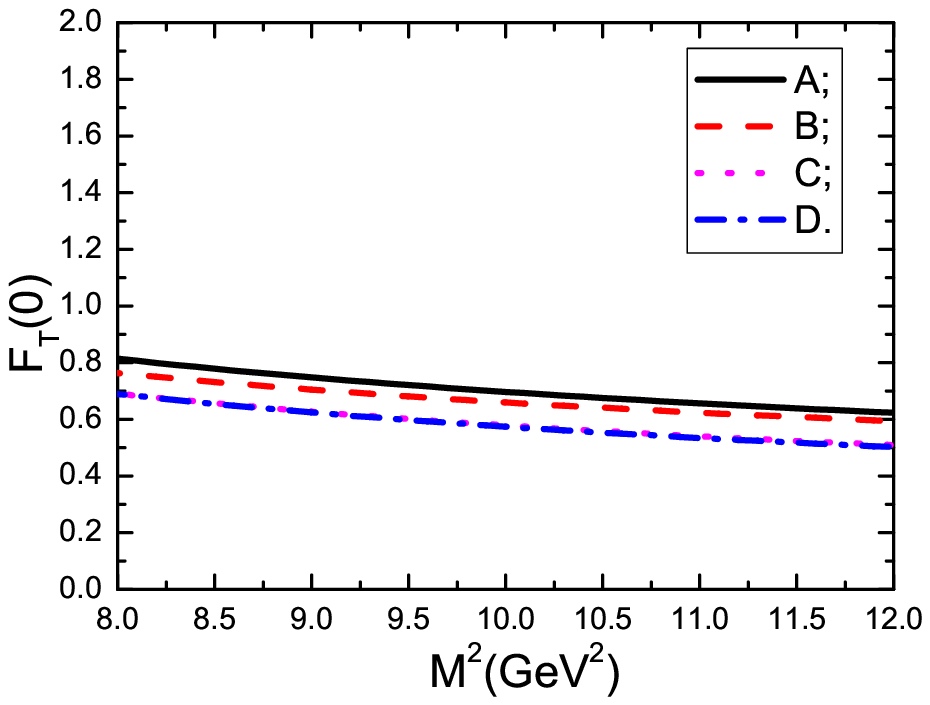}
    \caption{The central values of the form-factors $F_{+}(0)$, $F_{-}(0)$ and $F_T(0)$ with variations of the Borel parameter $M^2$, where $A$, $B$, $C$ and $D$ denote
    the transitions $B-a_0(1450)$, $B-K_0^*(1430)$, $B_s-K_0^*(1430)$ and $B_s-f_0(1500)$, respectively. }
\end{figure}

We take into account all uncertainties of the input parameters and
  obtain the numerical values of the transitions form-factors, which are  shown explicitly in Figs.4-5.
\begin{figure}
 \centering
 \includegraphics[totalheight=4cm,width=4.5cm]{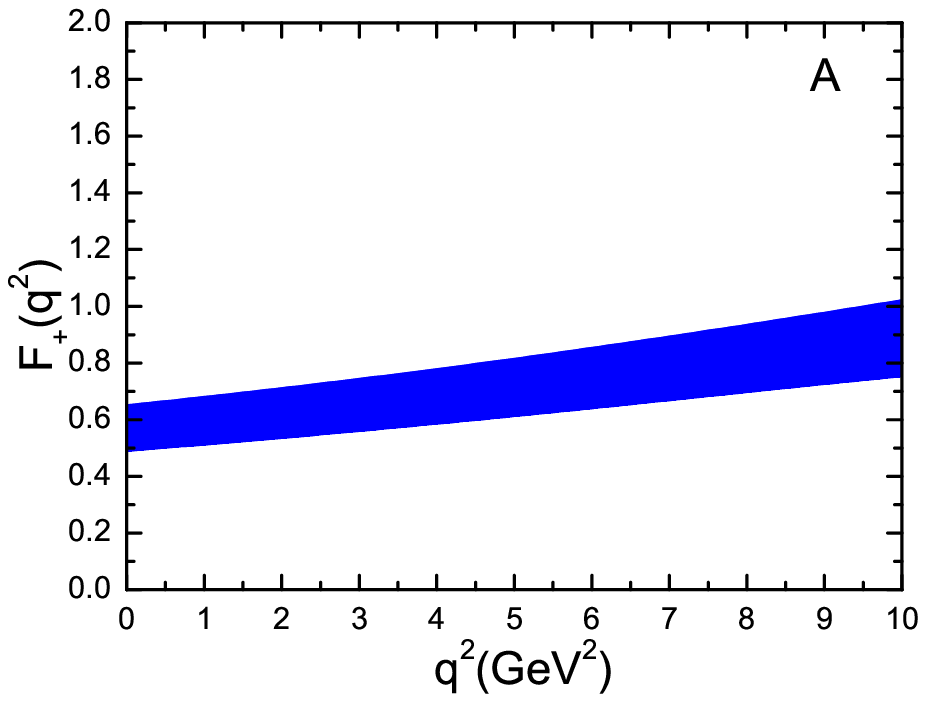}
 \includegraphics[totalheight=4cm,width=4.5cm]{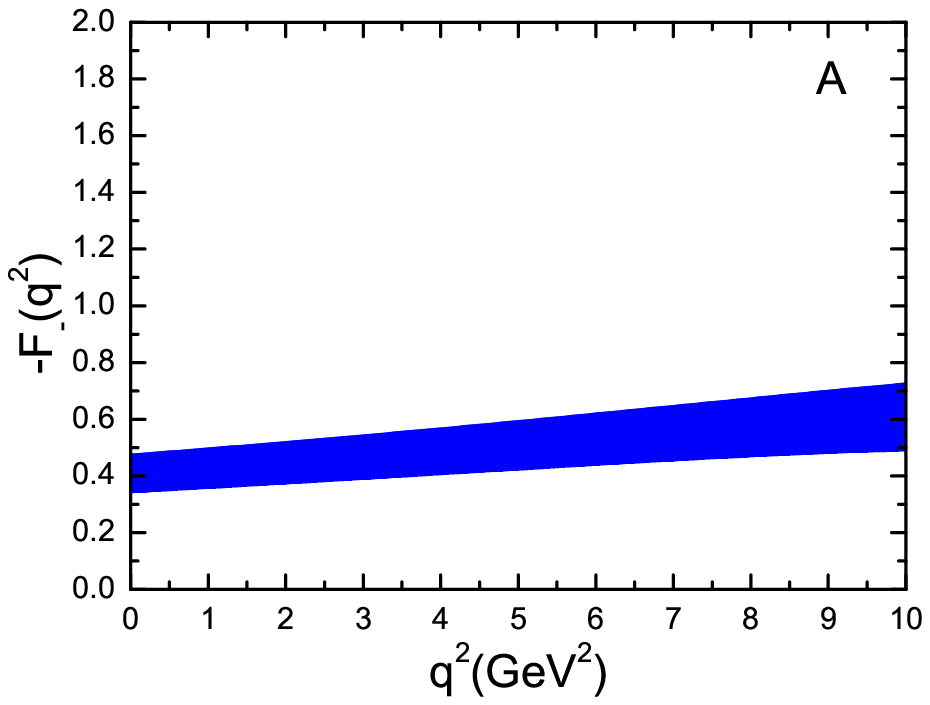}
 \includegraphics[totalheight=4cm,width=4.5cm]{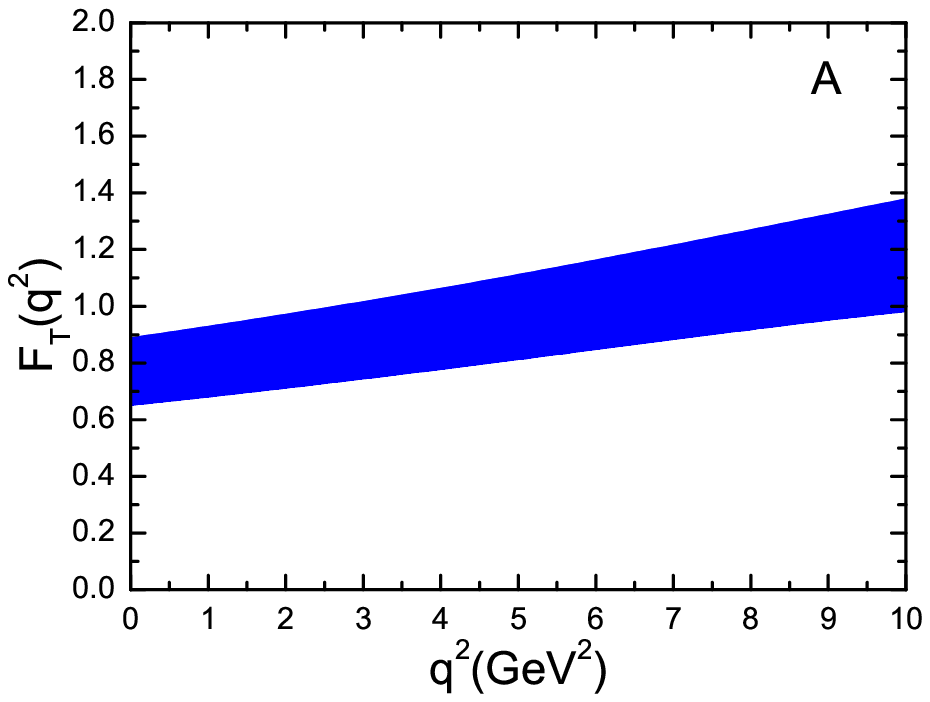}
  \includegraphics[totalheight=4cm,width=4.5cm]{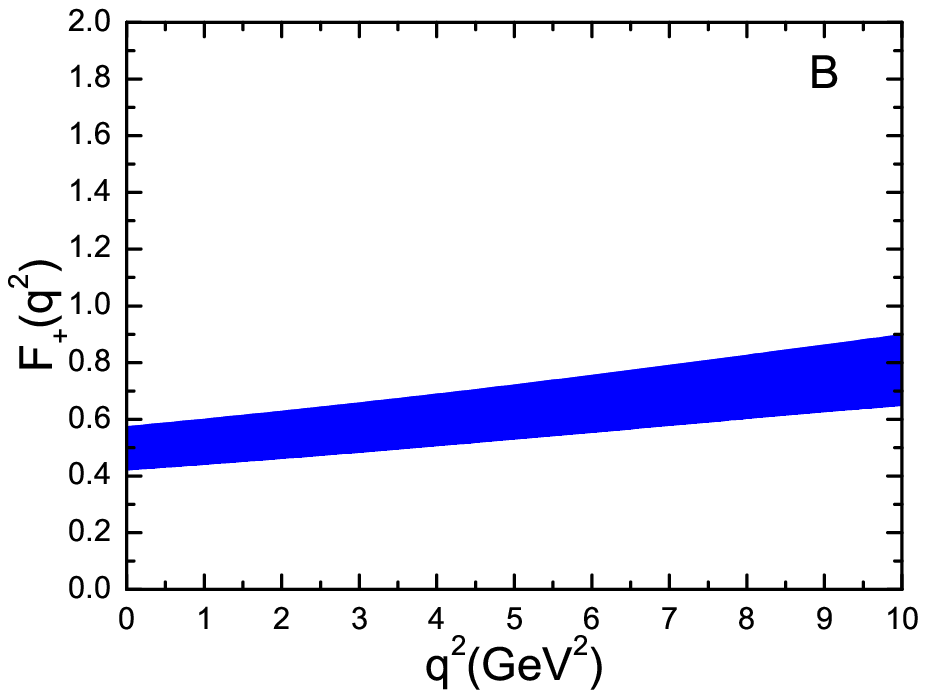}
 \includegraphics[totalheight=4cm,width=4.5cm]{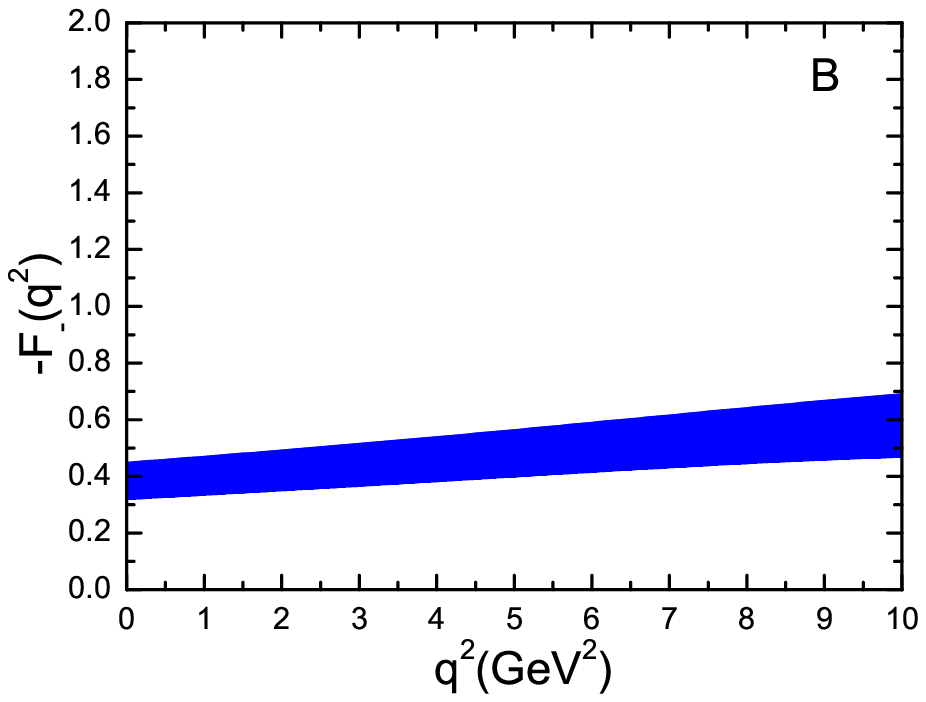}
  \includegraphics[totalheight=4cm,width=4.5cm]{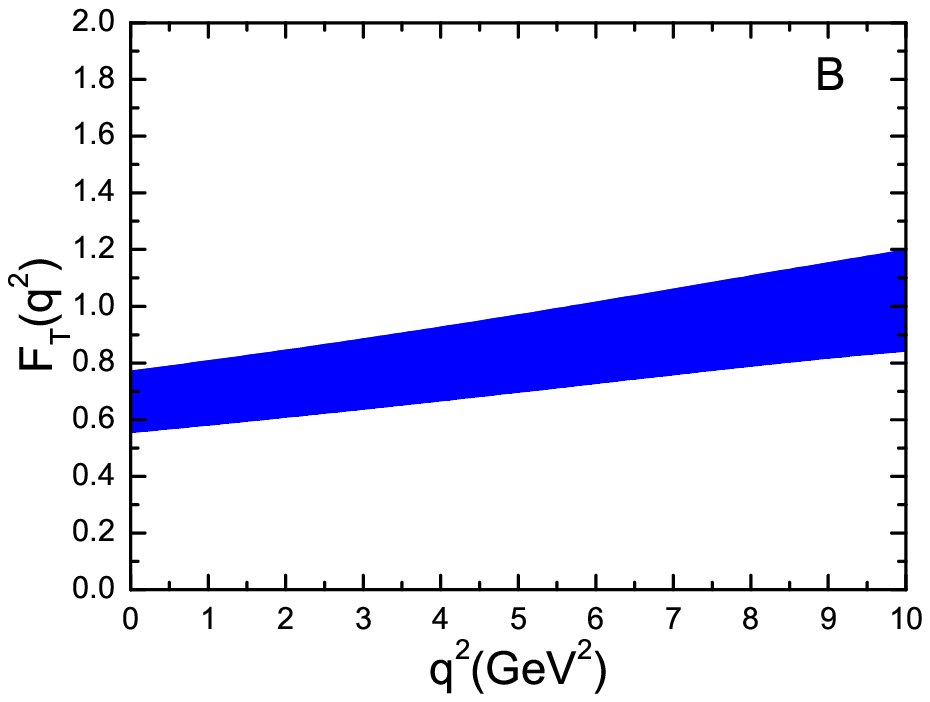}
   \includegraphics[totalheight=4cm,width=4.5cm]{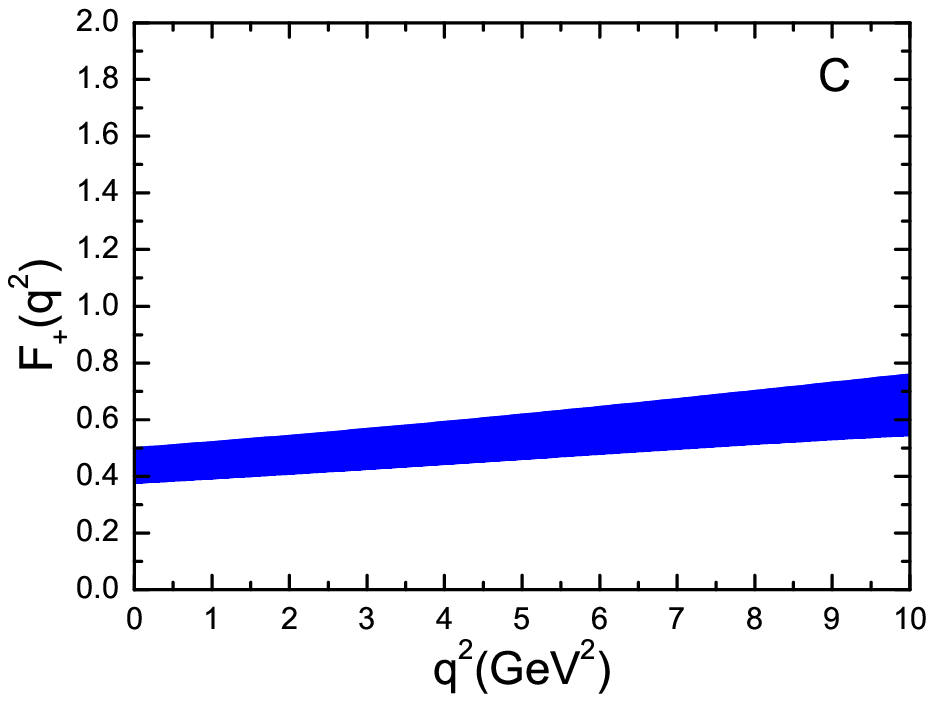}
 \includegraphics[totalheight=4cm,width=4.5cm]{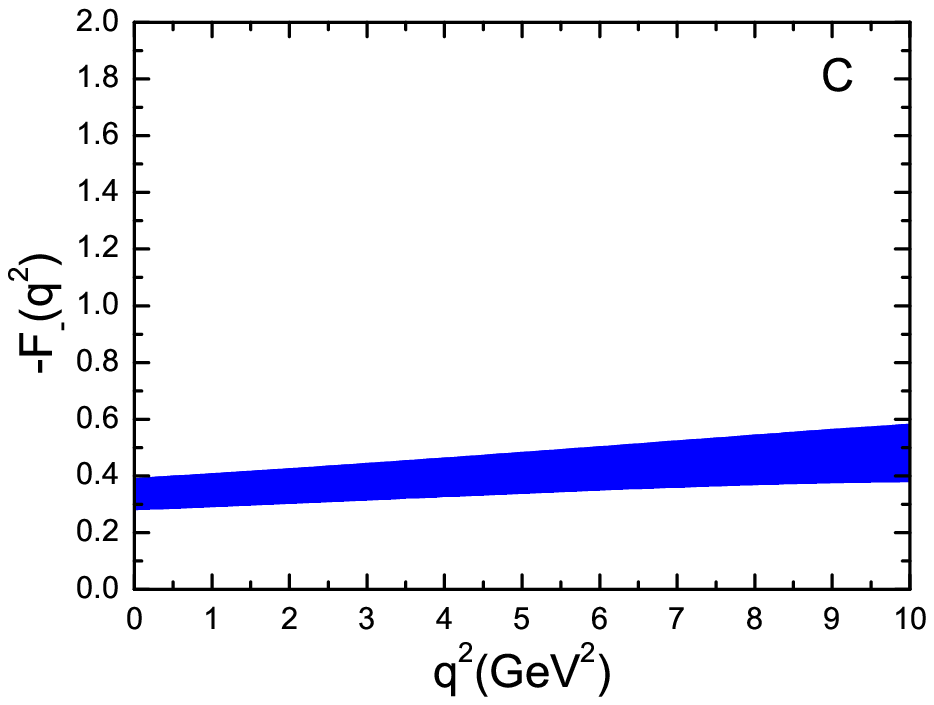}
  \includegraphics[totalheight=4cm,width=4.5cm]{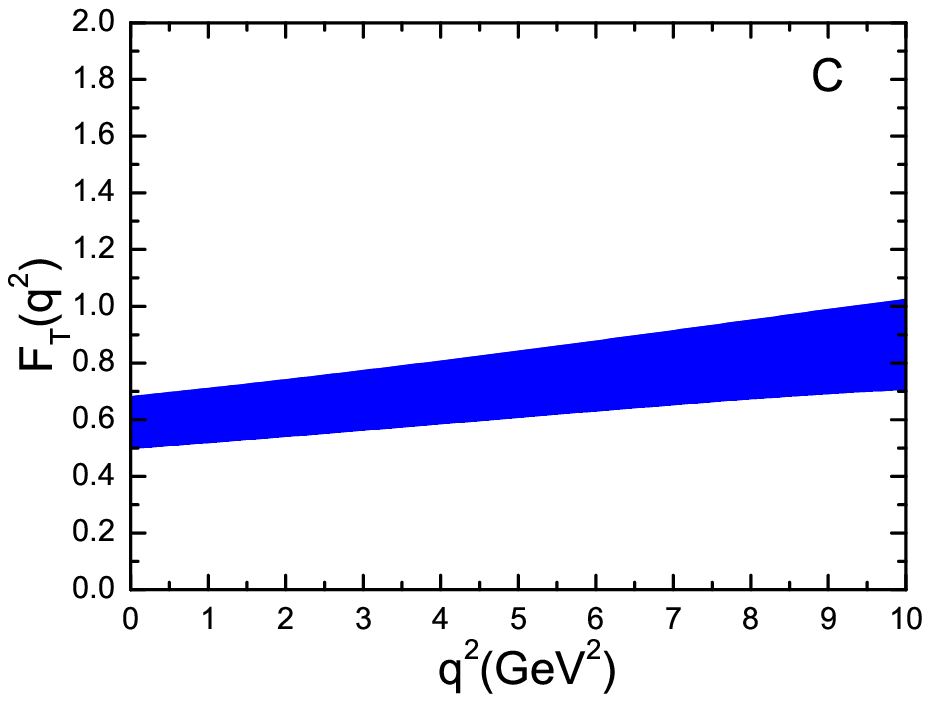}
  \includegraphics[totalheight=4cm,width=4.5cm]{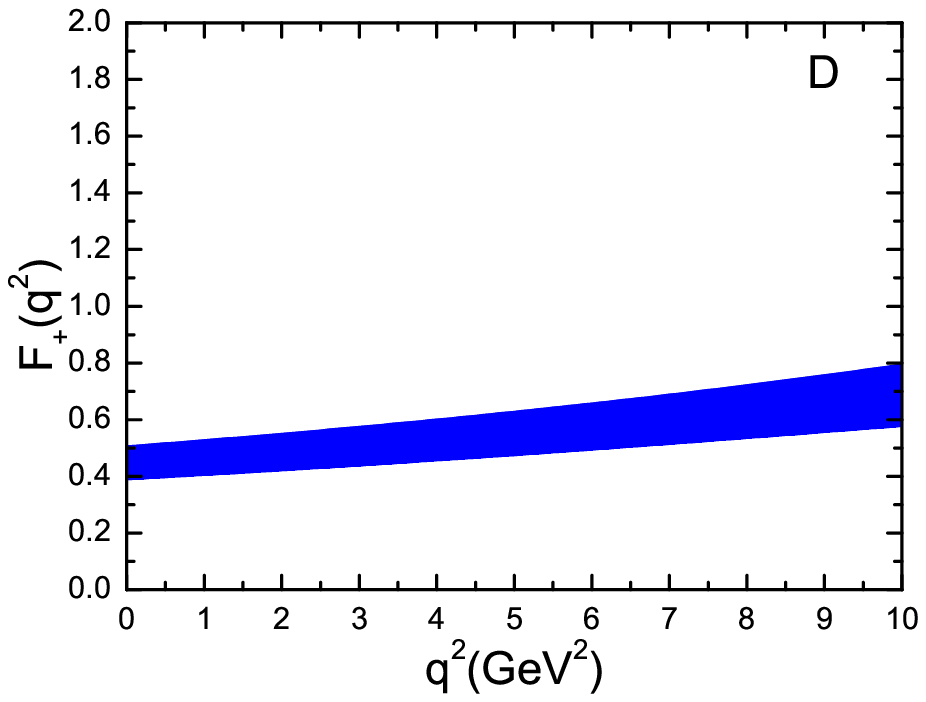}
 \includegraphics[totalheight=4cm,width=4.5cm]{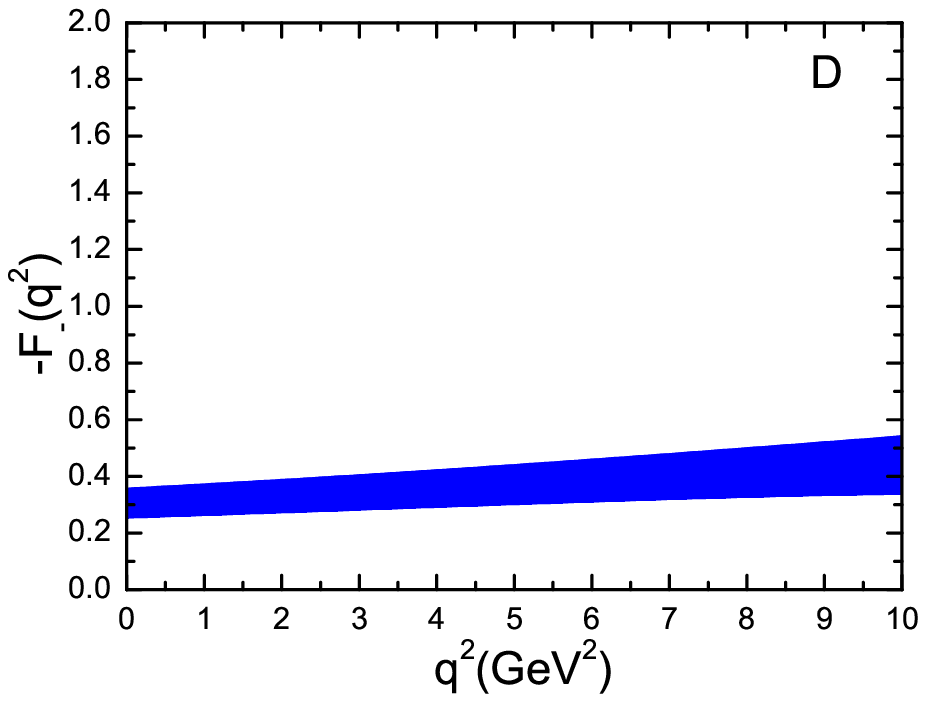}
  \includegraphics[totalheight=4cm,width=4.5cm]{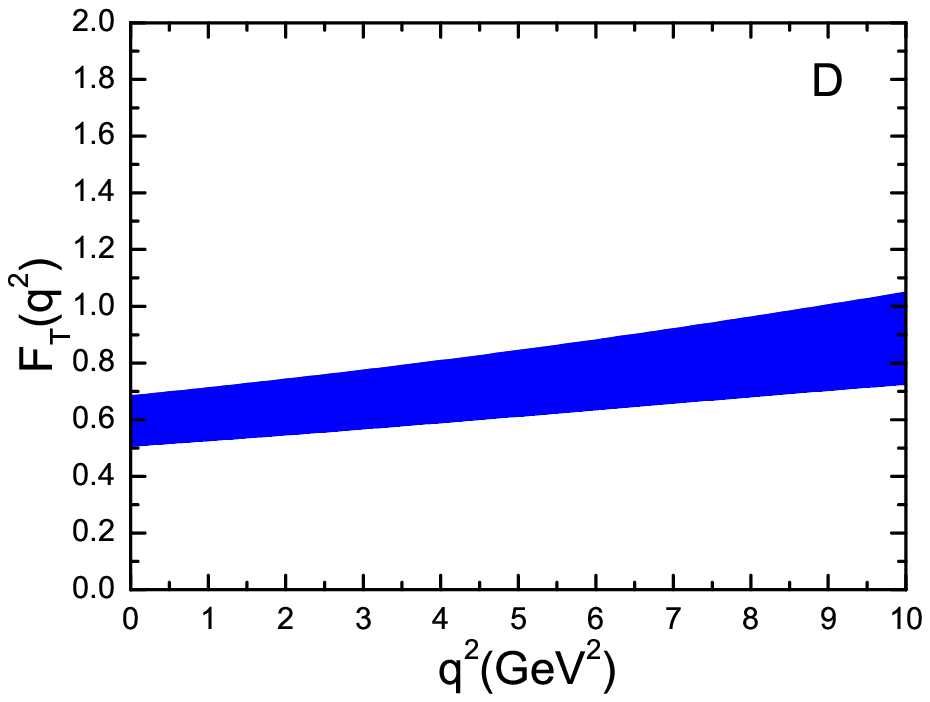}
    \caption{The  form-factors $F_{+}(q^2)$, $F_{-}(q^2)$ and $F_T(q^2)$ with variations of the momentum transfer squared $q^2$, where $A$, $B$, $C$ and $D$ denote
    the transitions $B-a_0(980)$, $B-\kappa(800)$, $B_s-\kappa(800)$ and $B_s-f_0(980)$, respectively.}
\end{figure}
\begin{figure}
 \centering
 \includegraphics[totalheight=4cm,width=4.8cm]{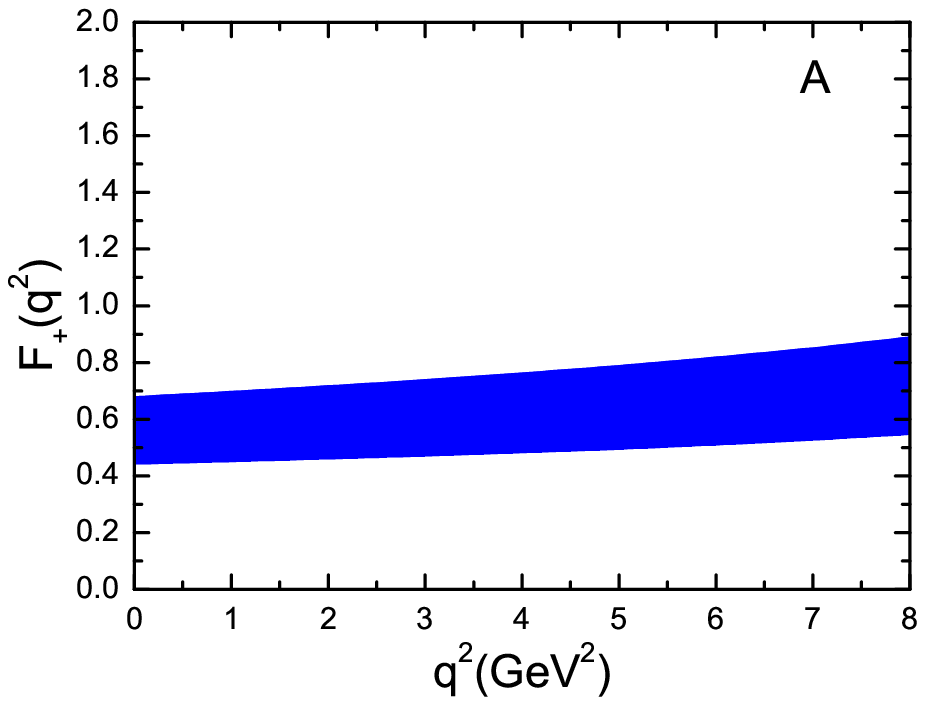}
 \includegraphics[totalheight=4cm,width=4.8cm]{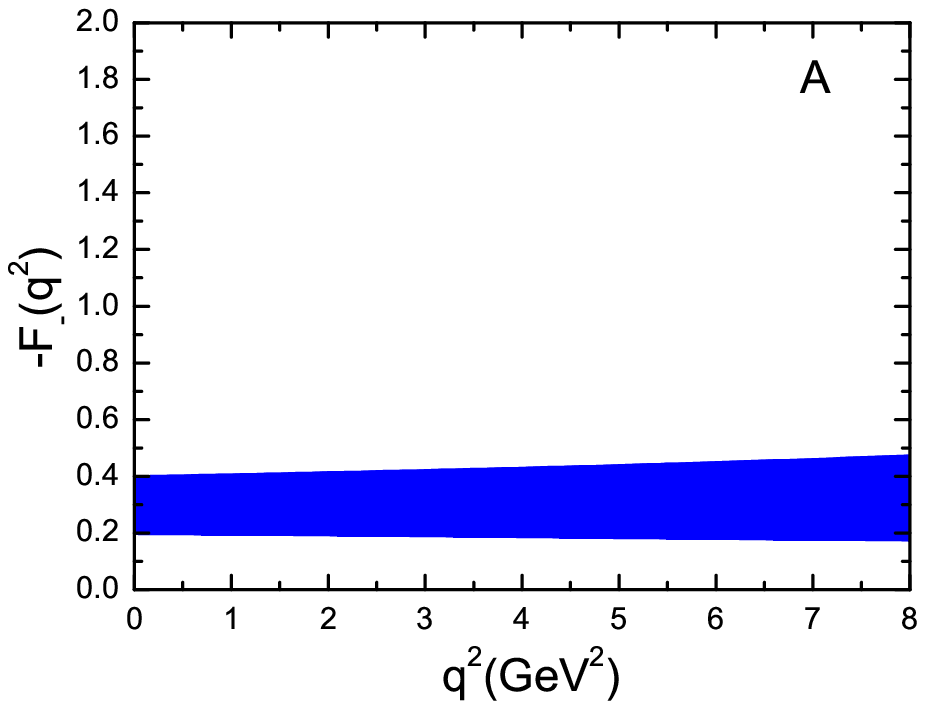}
  \includegraphics[totalheight=4cm,width=4.8cm]{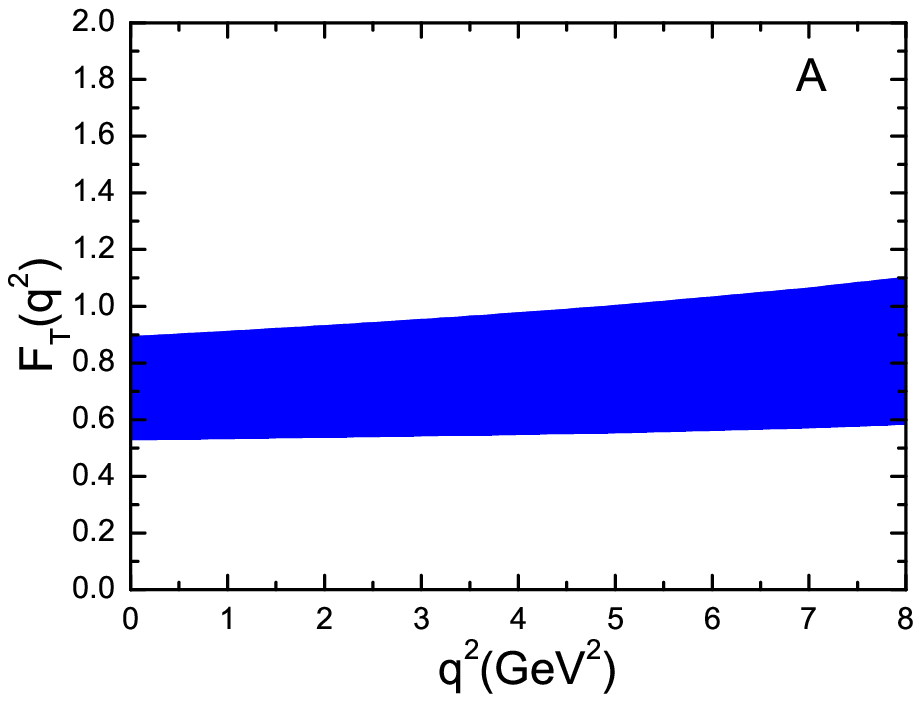}
  \includegraphics[totalheight=4cm,width=4.8cm]{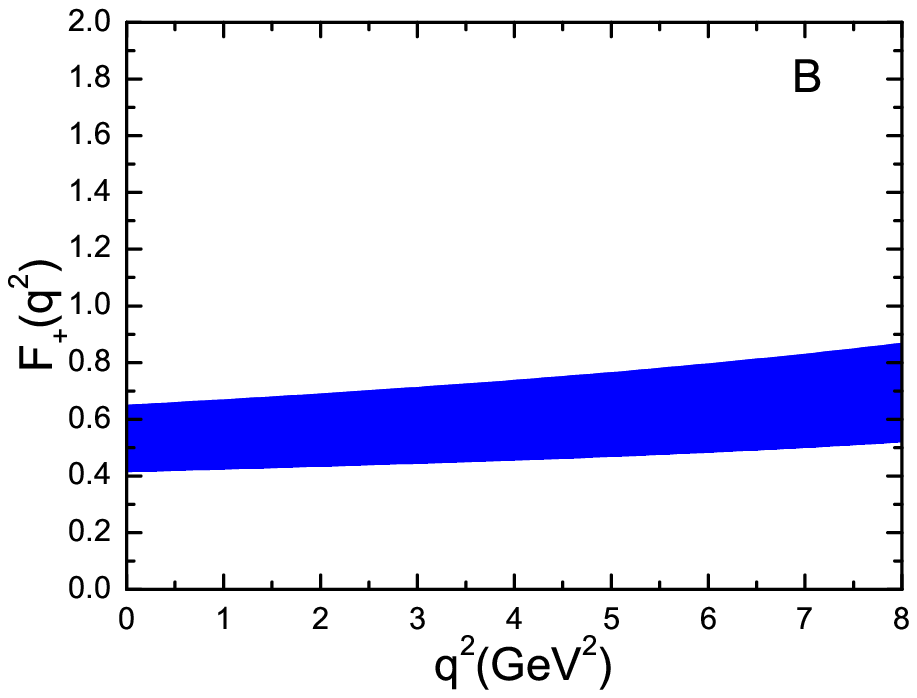}
 \includegraphics[totalheight=4cm,width=4.8cm]{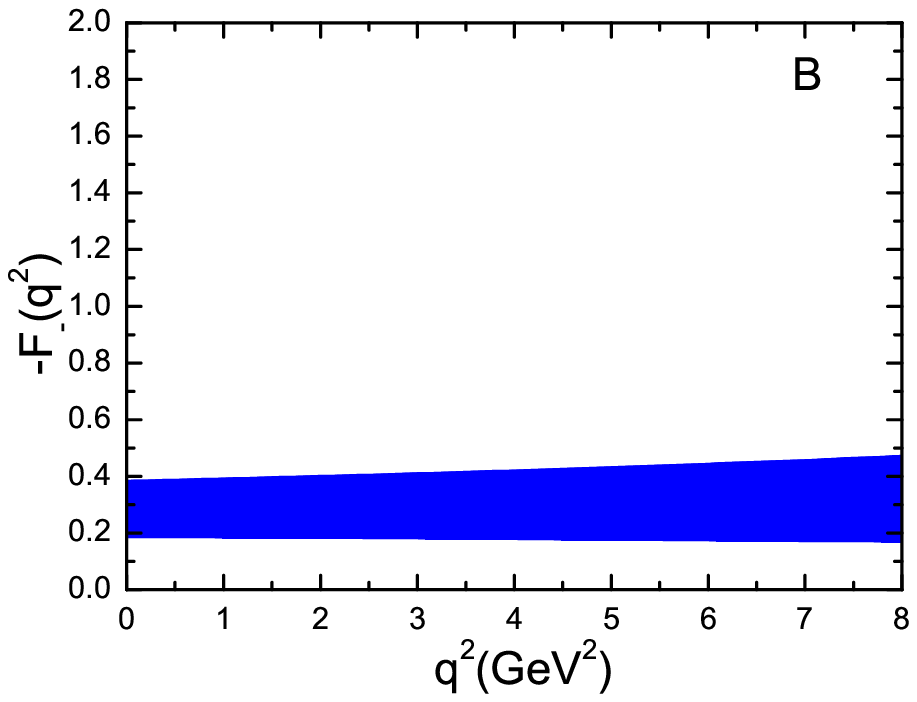}
  \includegraphics[totalheight=4cm,width=4.8cm]{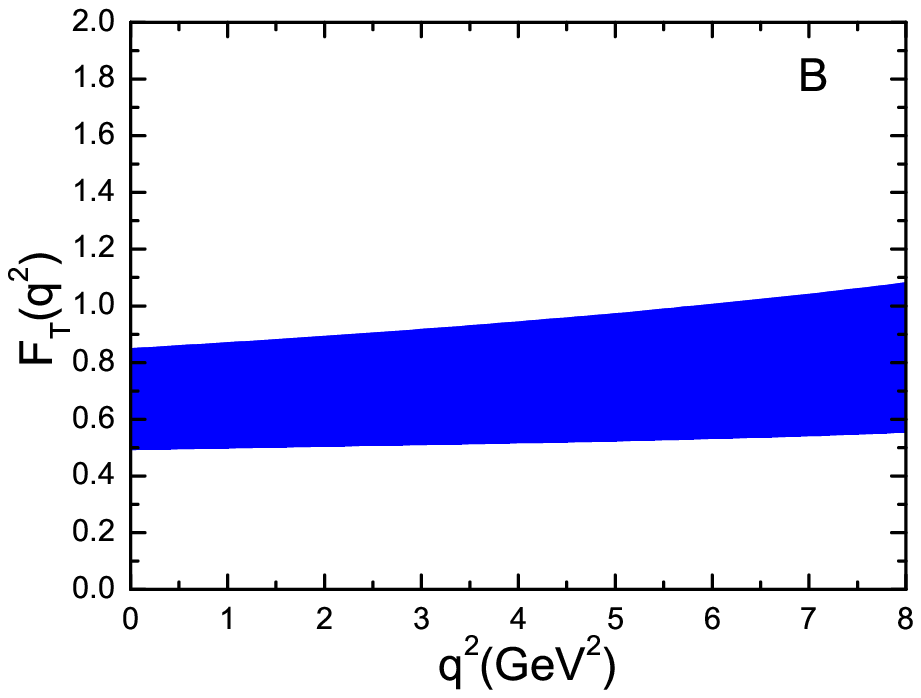}
  \includegraphics[totalheight=4cm,width=4.8cm]{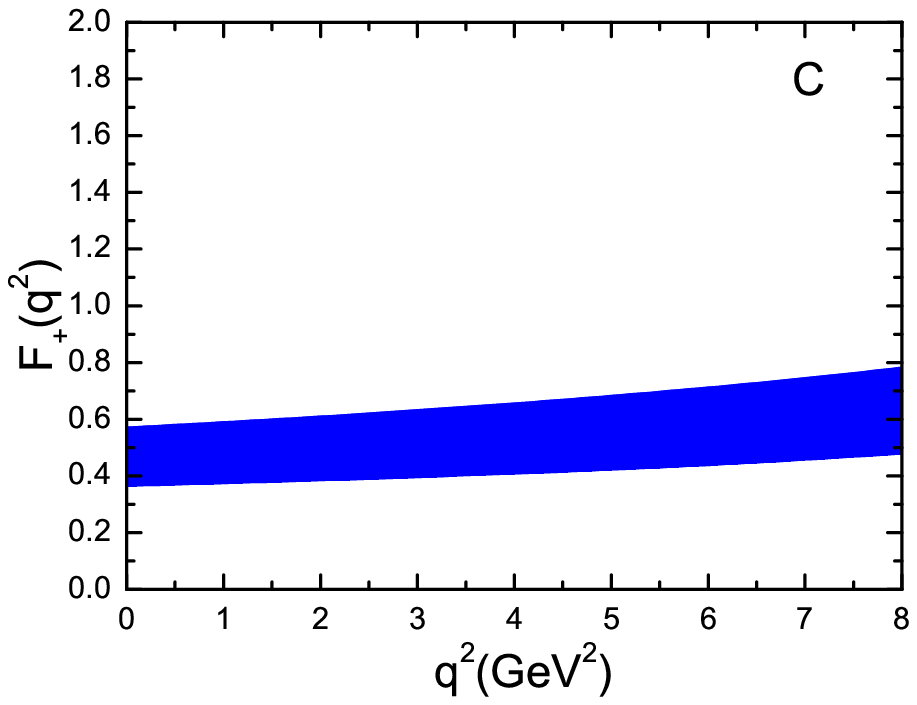}
 \includegraphics[totalheight=4cm,width=4.8cm]{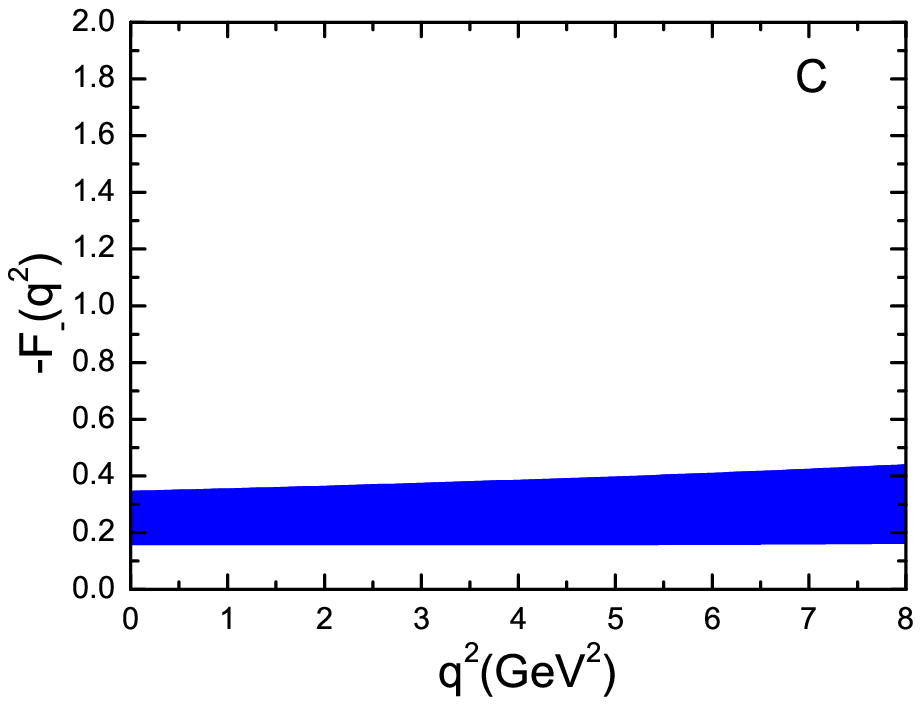}
  \includegraphics[totalheight=4cm,width=4.8cm]{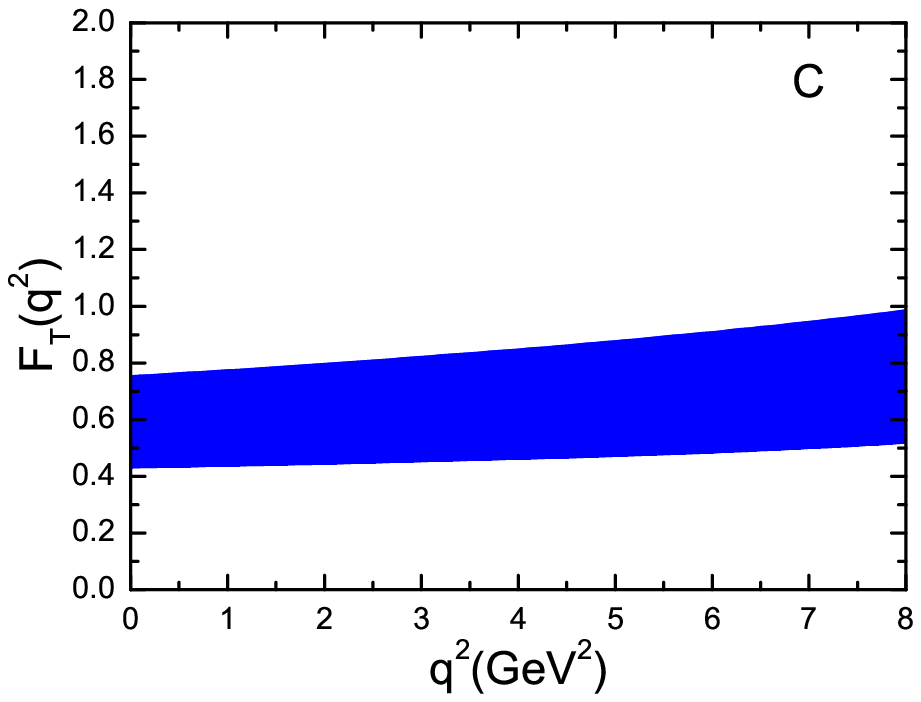}
  \includegraphics[totalheight=4cm,width=4.8cm]{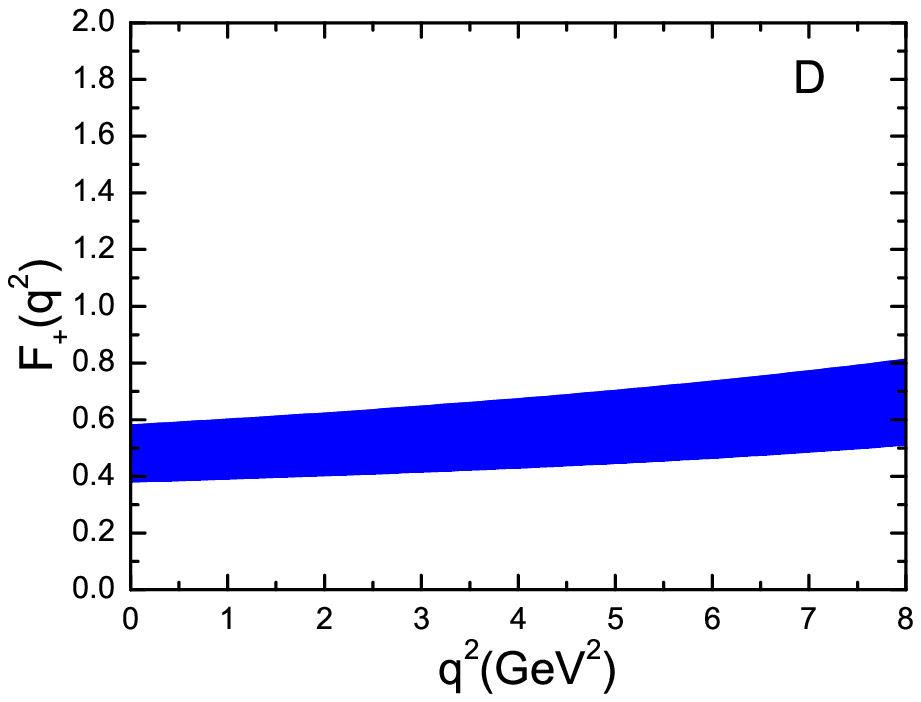}
 \includegraphics[totalheight=4cm,width=4.8cm]{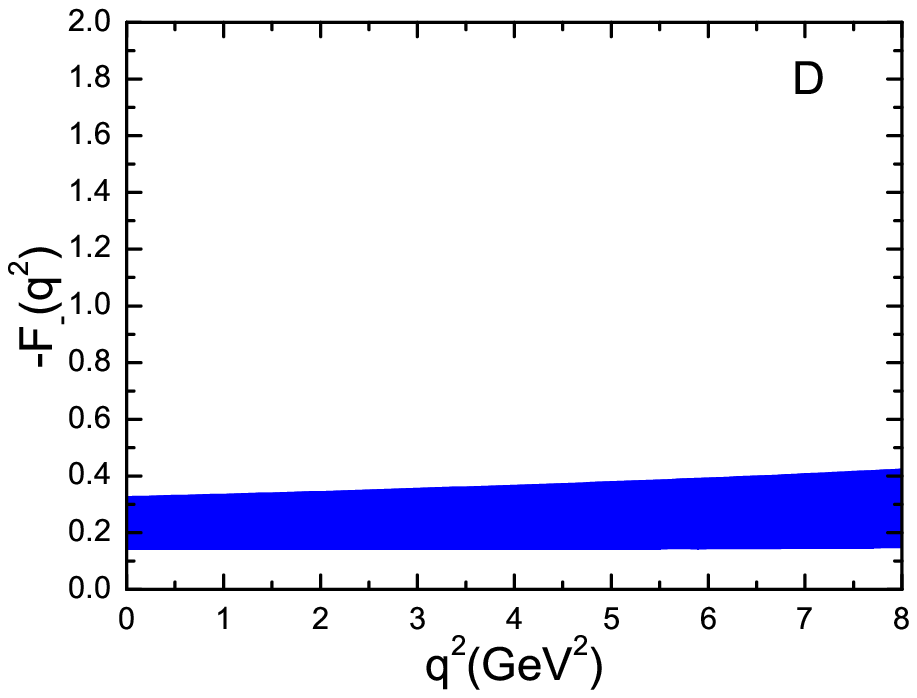}
 \includegraphics[totalheight=4cm,width=4.8cm]{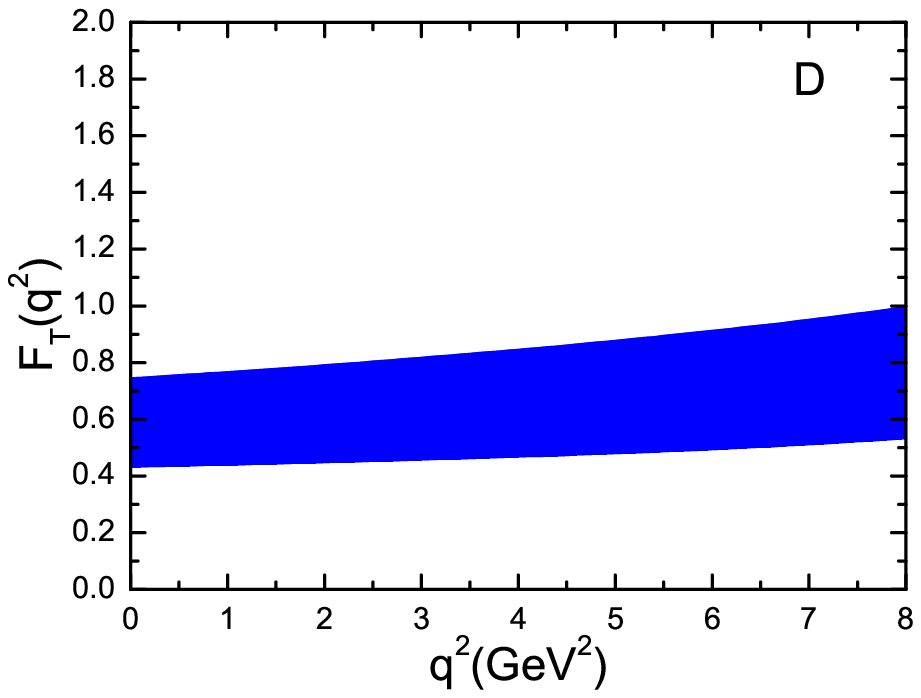}
    \caption{The  form-factors $F_{+}(q^2)$, $F_{-}(q^2)$ and $F_T(q^2)$ with variations of the momentum transfer squared  $q^2$, where $A$, $B$, $C$ and $D$ denote
    the transitions $B-a_0(1450)$, $B-K_0^*(1430)$, $B_s-K_0^*(1430)$ and $B_s-f_0(1500)$, respectively.}
\end{figure}

 The numerical values of the form-factors are fitted into the single-pole form,
 \begin{eqnarray}
 F_{i}(q^2)&=&\frac{F_{i}(0)}{1-a_{i}\frac{q^2}{m_B^2}} \, ,
 \end{eqnarray}
with the MINUIT, where $m_B=5.28\,\rm{GeV}$, $i=+,-,T$, the values of the $F_{i}(0)$ and $a_{i}$ are shown explicitly in Tables 1-3.
The form-factors can be taken as basic input parameters, and  study the semi-leptonic and non-leptonic $B$-decays  to the scalar mesons   so as to  shed light on the structures of the scalar mesons or explore  the CKM matrix elements.

In Tables 4-7, we present the $B-S$ transition form-factors from the LCSR \cite{WYM08,WW10,SYJ11,WXG13}, the three-point QCD sum rules  \cite{YMZ06,Alv07,GK09}, and also  the results  without perturbative ${\mathcal{O}}(\alpha_s)$ corrections in the LCSR. From the Table, we  can see that the perturbative ${\mathcal{O}}(\alpha_s)$ corrections are about $(15-35)\%$ and $(5-20)\%$ for the $B-S$ form-factors with the scalar mesons bellow and above 1 GeV, respectively, we should take them into account consistently.

 \begin{table}
\begin{center}
\begin{tabular}{|c|c|c|c|c|c|c|c|c|}\hline\hline
                       & $F_{+}(0)$        & $a_{+}$                   \\ \hline
  $B-a_0(980) $        & $0.576\pm0.042$   & $0.987\pm0.251$             \\ \hline
  $B-\kappa(800)$      & $0.504\pm0.039$   & $0.988\pm0.266$                  \\ \hline
  $B_s-\kappa(800)$    & $0.442\pm0.033$   & $0.904\pm0.274$                  \\ \hline
  $B_s-f_0(980) $      & $0.448\pm0.032$   & $0.952\pm0.257$                 \\ \hline
  $B-a_0(1450) $       & $0.549\pm0.071$   & $0.743\pm0.656$                 \\ \hline
  $B-K_0^*(1430)$      & $0.523\pm0.070$   & $0.795\pm0.669$                    \\ \hline
  $B_s-K_0^*(1430)$    & $0.458\pm0.062$   & $0.885\pm0.644$                    \\ \hline
  $B_s-f_0(1500) $     & $0.470\pm0.059$   & $0.941\pm0.595$                  \\ \hline
         \hline
\end{tabular}
\end{center}
\caption{ The parameters of the fitted  transition form-factors $F_{+}(q^2)$.    }
\end{table}

\begin{table}
\begin{center}
\begin{tabular}{|c|c|c|c|c|c|c|c|c|}\hline\hline
                           & $-F_{-}(0)$       & $a_{-}$          \\ \hline
  $B-a_0(980) $            & $0.414\pm0.036$   & $0.904\pm0.319$    \\ \hline
  $B-\kappa(800)$          & $0.390\pm0.034$   & $0.934\pm0.314$         \\ \hline
  $B_s-\kappa(800)$        & $0.340\pm0.030$   & $0.829\pm0.342$         \\ \hline
  $B_s-f_0(980) $          & $0.305\pm0.029$   & $0.830\pm0.377$        \\ \hline
  $B-a_0(1450) $           & $0.287\pm0.067$   & $0.190\pm1.445$        \\ \hline
  $B-K_0^*(1430)$          & $0.275\pm0.064$   & $0.330\pm1.402$           \\ \hline
  $B_s-K_0^*(1430)$        & $0.240\pm0.058$   & $0.518\pm1.353$           \\ \hline
  $B_s-f_0(1500) $         & $0.222\pm0.057$   & $0.565\pm1.418$         \\ \hline
         \hline
\end{tabular}
\end{center}
\caption{ The parameters of the fitted  transition form-factors $F_{-}(q^2)$.    }
\end{table}

 \begin{table}
\begin{center}
\begin{tabular}{|c|c|c|c|c|c|c|}\hline\hline
                       & $F_{T}(0)$        & $a_{T}$                  \\ \hline
  $B-a_0(980) $        & $0.778\pm0.062$   & $0.961\pm0.278$            \\ \hline
  $B-\kappa(800)$      & $0.673\pm0.056$   & $0.970\pm0.288$                 \\ \hline
  $B_s-\kappa(800)$    & $0.596\pm0.049$   & $0.877\pm0.304$                 \\ \hline
  $B_s-f_0(980) $      & $0.596\pm0.048$   & $0.900\pm0.299$                \\ \hline
  $B-a_0(1450) $       & $0.693\pm0.112$   & $0.511\pm0.893$                \\ \hline
  $B-K_0^*(1430)$      & $0.657\pm0.109$   & $0.598\pm0.900$                   \\ \hline
  $B_s-K_0^*(1430)$    & $0.575\pm0.098$   & $0.718\pm0.874$                   \\ \hline
  $B_s-f_0(1500) $     & $0.570\pm0.095$   & $0.778\pm0.835$                 \\ \hline
         \hline
\end{tabular}
\end{center}
\caption{ The parameters of the fitted  transition form-factors $F_{T}(q^2)$.    }
\end{table}

\begin{table}
\begin{center}
\begin{tabular}{|c|c|c|c|c|c|c|}\hline\hline
                 &$B-a_0(980)$          &$B-\kappa(800)$    &$B_s-\kappa(800)$    &$B_s-f_0(980)$          \\ \hline
  \cite{WW10}    &                      &                   &                     &$0.19/--$              \\ \hline
  \cite{SYJ11}   &$0.56/0.56$           &$0.46/0.46$        &$0.53/0.53$          &$0.44/0.44$            \\ \hline
 \cite{GK09}     &                      &                   &                     &$0.12/0.17$            \\ \hline
   This work     &$0.58/0.41$           &$0.50/0.39$        &$0.44/0.34$          &$0.45/0.31$            \\ \hline
   LO            &$0.45/0.32$           &$0.39/0.30$        &$0.36/0.29$          &$0.38/0.25$            \\ \hline
  \hline
\end{tabular}
\end{center}
\caption{ The values of the $F_{+}(0)/\left( -F_{-}(0)\right)$ of the transitions $B-a_0(980)$,
          $B-\kappa(800)$, $B_s-\kappa(800)$ and    $B_s-f_0(980)$  from the LCSR and QCD sum rules, where the LO denotes the leading order contributions. }
\end{table}

\begin{table}
\begin{center}
\begin{tabular}{|c|c|c|c|c|c|c|}\hline\hline
                      &$B-a_0(1450)$         &$B-K_0^*(1430)$    &$B_s-K_0^*(1430)$   &$B_s-f_0(1500)$  \\ \hline
  \cite{WYM08}        &$0.52/0.44$           &$0.49/0.41$        &$0.42/0.34$         &$0.43/0.37$        \\ \hline
  \cite{SYJ11}        &$0.53/0.53$           &$0.49/0.49$        &$0.44/0.44$         &$0.41/0.41$     \\ \hline
  \cite{WXG13}        &$0.44/0.26$           &$0.45/0.28$        &$0.39/0.25$         &$0.38/0.24$     \\ \hline
   \cite{YMZ06}       &                      &                   &$0.24/--$           &                 \\ \hline
  \cite{Alv07}        &                      &$0.31/0.31$        &                    &                 \\ \hline
   This work          &$0.55/0.29$           &$0.52/0.28$        &$0.46/0.24$         &$0.47/0.22$     \\ \hline
   LO                 &$0.51/0.26$           &$0.48/0.25$        &$0.43/0.22$         &$0.44/0.21$     \\ \hline
  \hline
\end{tabular}
\end{center}
\caption{ The values of the $F_{+}(0)/\left( -F_{-}(0)\right)$ of the transitions $B-a_0(1450)$,
         $B-K_0^*(1430)$, $B_s-K_0^*(1430)$ and  $B_s-f_0(1500)$  from the LCSR and QCD sum rules, where the LO denotes the leading order contributions. }
\end{table}

\begin{table}
\begin{center}
\begin{tabular}{|c|c|c|c|c|c|c|}\hline\hline
                 &$B-a_0(980)$          &$B-\kappa(800)$     &$B_s-\kappa(800)$  &$B_s-f_0(980)$          \\ \hline
  \cite{WW10}    &                      &                    &                   &$0.23$              \\ \hline
  \cite{SYJ11}   &                      &$0.58$              &                   &$0.58 $            \\ \hline
   This work     &$0.78$                &$0.67$              &$0.60$             &$0.60$            \\ \hline
   LO            &$0.51$                &$0.44$              &$0.42$             &$0.43$            \\ \hline
  \hline
\end{tabular}
\end{center}
\caption{ The values of the $F_{T}(0)$ of the transitions $B-a_0(980)$,
          $B-\kappa(800)$, $B_s-\kappa(800)$ and     $B_s-f_0(980)$ from the LCSR and QCD sum rules, where the LO denotes the leading order contributions. }
\end{table}

\begin{table}
\begin{center}
\begin{tabular}{|c|c|c|c|c|c|c|}\hline\hline
                      &$B-a_0(1450)$         &$B-K_0^*(1430)$     &$B_s-K_0^*(1430)$   &$B_s-f_0(1500)$  \\ \hline
  \cite{WYM08}        &$0.66$                &$0.60$              &$0.52$              &$0.56$        \\ \hline
  \cite{SYJ11}        &                      &$0.69$              &                    &$0.59$           \\ \hline
  \cite{WXG13}        &$0.43$                &$0.46$              &$0.41$              &$0.40 $            \\ \hline
  \cite{Alv07}        &                      &$0.26$              &                    &                 \\ \hline
   This work          &$0.69$                &$0.66$              &$0.58$              &$0.57$     \\ \hline
   LO                 &$0.55$                &$0.51$              &$0.47$              &$0.47$     \\ \hline
  \hline
\end{tabular}
\end{center}
\caption{ The values of the $F_{T}(0)$ of the transitions $B-a_0(1450)$,
         $B-K_0^*(1430)$, $B_s-K_0^*(1430)$ and   $B_s-f_0(1500)$ from the LCSR and QCD sum rules, where the LO denotes the leading order contributions. }
\end{table}

\section{Conclusion}
In the article,  we assume  the two scalar nonet mesons below and above 1 GeV are all $\bar{q}q$ states,
  in case I, the scalar mesons below 1 GeV are the ground states, in case II, the scalar mesons above 1 GeV are the ground states.
   We calculate the $B-S$ transition form-factors by taking into account the perturbative  ${\mathcal{O}}(\alpha_s)$
 corrections to the twist-2 terms  using the LCSR and fit the numerical values of the form-factors into the single-pole forms.
 The numerical results indicate that the perturbative  ${\mathcal{O}}(\alpha_s)$ corrections are  about $(15-35)\%$ and $(5-20)\%$ for the $B-S$ form-factors with the scalar mesons bellow and above 1 GeV, respectively, we should take them into account consistently.
The form-factors can be taken as basic input parameters, and  study the semi-leptonic and non-leptonic $B$-decays to the scalar mesons   so as to  shed light on the structures of the scalar mesons or explore  the CKM matrix elements.

\section*{Acknowledgements}
This  work is supported by National Natural Science Foundation,
Grant Numbers 11375063, and Natural Science Foundation of Hebei province, Grant Number A2014502017.

\end{document}